\documentclass[
reprint,twocolumn,nofootinbib,floatfix,
amsmath,amssymb,aps,prd,superscriptaddress]{revtex4-2}

\usepackage{xspace}
\usepackage{aas_macros}
\usepackage{graphicx}% Include figure files
\usepackage{graphics}
\usepackage[table,svgnames,dvipsnames]{xcolor}
\usepackage{hyperref}% add hypertext capabilities
\hypersetup{colorlinks=true,citecolor=blue,filecolor=blue,urlcolor=blue,}
\usepackage{orcidlink} % for author list
\usepackage{hanging} % for affiliations

\def\be{\begin{equation}}
\def\ee{\end{equation}}

\def\ba#1\ea{\begin{align*}#1\end{align*}}

\defcitealias{dMdB2020}{\texttt{dMdB20}}

\defcitealias{DESI2024IV}{\texttt{DESI2024-IV}}
\newcommand{\DESIIV}{\citetalias{DESI2024IV}\xspace}

\newcommand{\lya}{Ly$\alpha$\xspace}
\newcommand{\lyaf}{Ly$\alpha$ forest\xspace}

\newcommand{\lyaxlya}{Ly$\alpha\times$Ly$\alpha$\xspace}
\newcommand{\lyaxlyaA}{Ly$\alpha$(A)$\times$Ly$\alpha$(A)\xspace}
\newcommand{\lyaxlyaB}{Ly$\alpha$(A)$\times$Ly$\alpha$(B)\xspace}

\newcommand{\lyaxqsoA}{Ly$\alpha$(A)$\times$QSO\xspace}
\newcommand{\lyaxqsoB}{Ly$\alpha$(B)$\times$QSO\xspace}

\newcommand{\hMpc}{h^{-1}\,\mathrm{Mpc}}
\newcommand{\apar}{\alpha_\parallel}
\newcommand{\aper}{\alpha_\perp}
\newcommand{\zeff}{z_{\rm eff}}

\usepackage[nameinlink,noabbrev]{cleveref}
\crefname{equation}{Eq.}{Eqs.}
\crefname{section}{Section}{Sections}
\crefname{figure}{Figure}{Figures}
\crefname{table}{Table}{Tables}
\crefname{appendix}{Appendix}{Appendices}
\Crefname{figure}{Figure}{Figures}
\Crefname{equation}{Equation}{Equations}
\Crefname{section}{Section}{Sections}
\Crefname{table}{Table}{Tables}

\begin{document}

\title{DESI DR2 Results I: Baryon Acoustic Oscillations from the Lyman Alpha Forest}

% Author list file generated with: mkauthlist 1.3.0+14.gcc6daf1.dirty 
% mkauthlist -f --sort --orcid -j revtex DESI-2024-0511_Mar14.csv DESI-2024-0511_Mar24.tex 
%% Orcid numbers may need \usepackage{orcidlink}.
%% Use \input to call the file

\author{M.~Abdul Karim\orcidlink{0009-0000-7133-142X}}
\affiliation{IRFU, CEA, Universit\'{e} Paris-Saclay, F-91191 Gif-sur-Yvette, France}

\author{J.~Aguilar}
\affiliation{Lawrence Berkeley National Laboratory, 1 Cyclotron Road, Berkeley, CA 94720, USA}

\author{S.~Ahlen\orcidlink{0000-0001-6098-7247}}
\affiliation{Physics Dept., Boston University, 590 Commonwealth Avenue, Boston, MA 02215, USA}

\author{C.~Allende~Prieto\orcidlink{0000-0002-0084-572X}}
\affiliation{Departamento de Astrof\'{\i}sica, Universidad de La Laguna (ULL), E-38206, La Laguna, Tenerife, Spain}
\affiliation{Instituto de Astrof\'{\i}sica de Canarias, C/ V\'{\i}a L\'{a}ctea, s/n, E-38205 La Laguna, Tenerife, Spain}

\author{O.~Alves}
\affiliation{Department of Physics, University of Michigan, Ann Arbor, MI 48109, USA}

\author{A.~Anand\orcidlink{0000-0003-2923-1585}}
\affiliation{Lawrence Berkeley National Laboratory, 1 Cyclotron Road, Berkeley, CA 94720, USA}

\author{U.~Andrade\orcidlink{0000-0002-4118-8236}}
\affiliation{Leinweber Center for Theoretical Physics, University of Michigan, 450 Church Street, Ann Arbor, Michigan 48109-1040, USA}
\affiliation{Department of Physics, University of Michigan, Ann Arbor, MI 48109, USA}

\author{E.~Armengaud\orcidlink{0000-0001-7600-5148}}
\affiliation{IRFU, CEA, Universit\'{e} Paris-Saclay, F-91191 Gif-sur-Yvette, France}

\author{A.~Aviles\orcidlink{0000-0001-5998-3986}}
\affiliation{Instituto de Ciencias F\'{\i}sicas, Universidad Nacional Aut\'onoma de M\'exico, Av. Universidad s/n, Cuernavaca, Morelos, C.~P.~62210, M\'exico}
\affiliation{Instituto Avanzado de Cosmolog\'{\i}a A.~C., San Marcos 11 - Atenas 202. Magdalena Contreras. Ciudad de M\'{e}xico C.~P.~10720, M\'{e}xico}

\author{S.~Bailey\orcidlink{0000-0003-4162-6619}}
\affiliation{Lawrence Berkeley National Laboratory, 1 Cyclotron Road, Berkeley, CA 94720, USA}

% ...

\collaboration{DESI Collaboration}
 %authorlist file
\email{spokespersons@desi.lbl.gov}

\begin{abstract}

We present the Baryon Acoustic Oscillation (BAO) measurements with the Lyman-$\alpha$ (\lya) forest from the second data release (DR2) of the Dark Energy Spectroscopic Instrument (DESI) survey. Our BAO measurements include both the auto-correlation of the \lya forest absorption observed in the spectra of high-redshift quasars and the cross-correlation of the absorption with the quasar positions. The total sample size is approximately a factor of two larger than the DR1 dataset, with forest measurements in over 820,000 quasar spectra and the positions of over 1.2 million quasars.  We describe several significant improvements to our analysis in this paper, and two supporting papers describe improvements to the synthetic datasets that we use for validation and how we identify damped \lya absorbers. Our main result is that we have measured the BAO scale with a statistical precision of 1.1\% along and 1.3\% transverse to the line of sight, for a combined precision of 0.65\% on the isotropic BAO scale at $\zeff = 2.33$. This excellent precision, combined with recent theoretical studies of the BAO shift due to nonlinear growth, motivated us to include a systematic error term in \lya BAO analysis for the first time. We measure the ratios $D_H(\zeff)/r_d = 8.632 \pm 0.098 \pm 0.026$ and $D_M(\zeff)/r_d = 38.99 \pm 0.52 \pm 0.12$, where $D_H = c/H(z)$ is the Hubble distance, $D_M$ is the transverse comoving distance, $r_d$ is the sound horizon at the drag epoch, and we quote both the statistical and the theoretical systematic uncertainty. The companion paper presents the BAO measurements at lower redshifts from the same dataset and the cosmological interpretation.
\end{abstract}

\maketitle

\section{Introduction} \label{sec:introduction}

The Baryon Acoustic Oscillations (BAO) scale is one of the most powerful and well understood tools for studying the expansion history and geometry of the Universe \citep[e.g.][]{Weinberg2013}. The BAO scale is due to primordial sound waves that were frozen into the matter distribution at the epoch of recombination, and they serve as a standard ruler that enables precise measurements of cosmic distances relative to the sound horizon at the drag epoch across a broad range of redshifts. When combined with observations of the Cosmic Microwave Background (CMB) and Type Ia supernovae, BAO measurements provide stringent tests of the standard cosmological model and its extensions \citep[e.g.][]{Planck2018,eBOSS2021,DESI2024.VI.KP7A,Pantheon2022,Union2023,DESY5SN}. 

The Lyman-$\alpha$ (\lya) forest is a unique window into the high-redshift Universe. Unlike the galaxies and quasars that are used to trace large-scale structure at lower redshifts, the \lya forest is observed as absorption features in the spectra of individual quasars that are produced by relatively low-density gas in the intergalactic medium \citep[e.g.][]{McQuinn2016}. The absorption traces large-scale structure through the relative amount of neutral hydrogen gas at a range of intervening redshifts. The \lya forest is unique because each sightline traces the matter distribution over a range of redshifts rather than at a single point, the forest absorbers are relatively less biased tracers of large scale structure compared to galaxies and quasars, and the forest samples the matter distribution over a broader range of spatial scales. These scales range from small scales that are dominated by gas pressure to the large scales that contain the BAO signal \citep{McDonald2003,McDonald2007,Slosar2011}. 

The \lya forest was developed into a tool to measure the BAO scale in large spectroscopic surveys with measurements of the forest auto-correlation \citep{Slosar2011} and then the cross-correlation of the forest with quasars \citep{FontRibera2013} with early data from the Baryon Oscillation Spectroscopic Survey \citep[BOSS,][]{Dawson2013}, a part of the Sloan Digital Sky Survey \citep[SDSS,][]{York2000,Eisenstein2011,Blanton2017}. Subsequent studies with additional data from BOSS \citep{Slosar2013,Busca2013,FontRibera2014,Delubac2015,Bautista2017} obtained the first detections of the BAO scale. These were followed by additional studies \citep{Blomqvist2019,deSainteAgathe2019} with data from the extended BOSS \citep[eBOSS,][]{Dawson2016} survey that substantially developed new methodologies for \lya forest analysis. These methodologies were needed to account for the dramatic improvements in the statistical precision of the \lya BAO measurements as the sample size grew from 14,000 quasars \citep{Slosar2011} to several hundred thousand. The final \lya sample from eBOSS measured the \lya forest auto-correlation with absorption spectra of over 210,000 quasars at $z > 2.1$ as well as the cross-correlation of the forest with the position of over 340,000 quasars at $z>1.77$ \citep{dMdB2020}. 

The Dark Energy Spectroscopic Instrument (DESI) represents a major step forward in the precise measurement of the BAO scale over a broad redshift range that includes the \lya forest \citep{DESI2016a.Science}. DESI is in the midst of a survey to measure precise redshifts for over 40 million galaxies and quasars in just five years, which is an order of magnitude larger than the final extragalactic sample from SDSS. This significant increase is possible because of DESI's highly efficient, massively multiplexed fiber spectrograph \citep{DESI2022.KP1.Instr}, which was designed to obtain 5000 spectra per observation. The DESI \lya forest dataset is already the largest ever obtained. Building on the 88,511 quasars \citep{Ramirez2024} in the DESI Early Data Release \cite{DESI2023b.KP1.EDR}, the DESI DR1 sample described in \cite{DESI2024.I.DR1} was used by  \cite[][hereafter \DESIIV]{DESI2024.IV.KP6} to analyze over 420,000 \lya forest spectra and their correlation with the spatial distribution of more than 700,000 quasars. In addition to the large sample, that work presented a number of further analysis improvements. These included a new series of tests of the analysis methodology prior to unblinding the results, as well as more tests with synthetic datasets, analysis of data splits, and tests of the robustness of the BAO measurements to reasonable alternative analysis choices. The final result measured $D_H(\zeff)/r_d$ with 2\% precision and $D_M(\zeff)/r_d$ with 2.4\% precision at $\zeff = 2.33$. 

This paper presents our measurement of the BAO scale with the first three years of DESI data, which we refer to as Data Release 2 \cite[DR2][]{DESI.DR2.DR2}. These observations were obtained between May 2021 and April 2024 and include over 820,000 \lya forest spectra and their correlation with the positions of over 1.2 million quasars. Our results are part of a comprehensive set of BAO measurements with the DESI DR2 dataset that span from the local universe to quasars at $z \sim 4.16$. The companion paper \citep{DESI.DR2.BAO.cosmo} presents the BAO measurements from galaxies and quasars at $z < 2$ and our cosmological interpretation of the measurements in both papers. That analysis includes constraints on dark energy models and potential extensions to $\Lambda$CDM.
There are also a series of supporting papers that provide more information about the analysis and explore additional implications \citep{Y3.clust-s1.Andrade.2025, Y3.lya-s2.Brodzeller.2025,Y3.lya-s1.Casas.2025,Y3.cpe-s1.Lodha.2025,Y3.cpe-s2.Elbers.2025}. 

The structure of this paper is as follows.  We start in \cref{sec:data} with a description of the DESI survey and the datasets used in our analysis. This includes the most relevant aspects of the spectroscopic analysis and redshift measurements, the generation of the QSO catalog and ancillary catalogs that provide information about the locations of Broad Absorption Lines and Damped Lyman-$\alpha$ absorption systems that can complicate the analysis, and the method we use to extract the \lyaf fluctuations from the spectra. In \cref{sec:correlations} we present a brief summary of how we measure the correlations, with an emphasis on changes from our previous work. The two largest changes are how we calculate the distortion matrix and how we model metal-line contamination. We present our measurements of the correlation function in \cref{sec:measurement}. This includes a discussion of the parameter choices for our baseline fit to the two-dimensional correlation functions, an analysis of the significance of outliers, and the systematic error budget. We conducted an extensive series of validation tests prior to unblinding our results. These tests with both mocks and data are described in \cref{sec:validation}. We discuss our results in \cref{sec:discussion}, including in the context of recent theoretical work on the potential shift of the BAO signal due to the non-linear growth of structure, as well as previous work. We provide our conclusions in the last section. Two appendices provide more information about the fit parameters and additional validation tests. 

\section{Data} \label{sec:data}

DESI was designed to study the nature of dark energy  with the first spectroscopic survey that met the criteria of a Stage-IV dark energy experiment \cite{DETF2006}, and to meet this goal in just five years of observations \citep{Snowmass2013.Levi,DESI2016b.Instr,DESI2016a.Science}.
The five-year plan is to measure 40 million galaxies and quasars from the local universe to beyond $z \sim 3.5$ over an area of 14,000 deg$^2$. This ambitious survey is feasible because the instrument was designed to obtain 5000 spectra per observation, the high throughput of the instrumentation combined with the 4-m aperture of the Mayall telescope at the Kitt Peak National Observatory, and the rapid reconfiguration time of the fiber positioner system. There is a paper that describes an overview of the instrumentation and the connection to the science requirements  \cite{DESI2022.KP1.Instr}. In addition, there are dedicated papers on the focal plane system \cite{FocalPlane.Silber.2023}, the corrector system \cite{Corrector.Miller.2023}, and the fiber system \cite{FiberSystem.Poppett.2024}. 

The DESI targets were selected from the significant imaging obtained for the DESI Legacy Imaging Surveys \cite{BASS.Zou.2017, LS.Overview.Dey.2019}. The quasar target selection algorithms are described in several papers \cite{QSOPrelim.Yeche.2020,QSO.TS.Chaussidon.2023} that are integrated into the target selection pipeline \cite{TS.Pipeline.Myers.2023}. We extensively tested the DESI survey operations and target selection during a period of survey validation \citep{DESI2023a.KP1.SV} prior to the start of the main survey in May 2021. The data from the survey validation period formed the early data release \citep[EDR,][]{DESI2023b.KP1.EDR}. We presented a number of papers at the time of this release that analyzed both the EDR data and the first two months of the main survey. These included numerous science results related to BAO with galaxies, quasars, and the \lya forest \citep{BAO.EDR.Moon.2023,Gordon2023}. The operation of the survey is described further in \cite{SurveyOps.Schlafly.2023}. 

Observations from the first year of the survey formed data release 1 \citep[DR1,][]{DESI2024.I.DR1}. That dataset forms the basis for a series of key science papers that include the measurement and validation of the two-point clustering of galaxies and quasars \citep{DESI2024.II.KP3}, BAO measurements of these galaxies and quasars \citep{DESI2024.III.KP4} and the \lya forest (\DESIIV), and the full-shape of correlation functions of the galaxies and quasars \citep{DESI2024.V.KP5}. There are also papers on the cosmological interpretation of these results, notably one focused on the BAO measurements \citep{DESI2024.VI.KP7A} and one that adds the full-shape information \citep{DESI2024.VII.KP7B}.  

\begin{figure*}
    \begin{minipage}{0.7\linewidth}
    \centering
    \includegraphics[width=\textwidth]{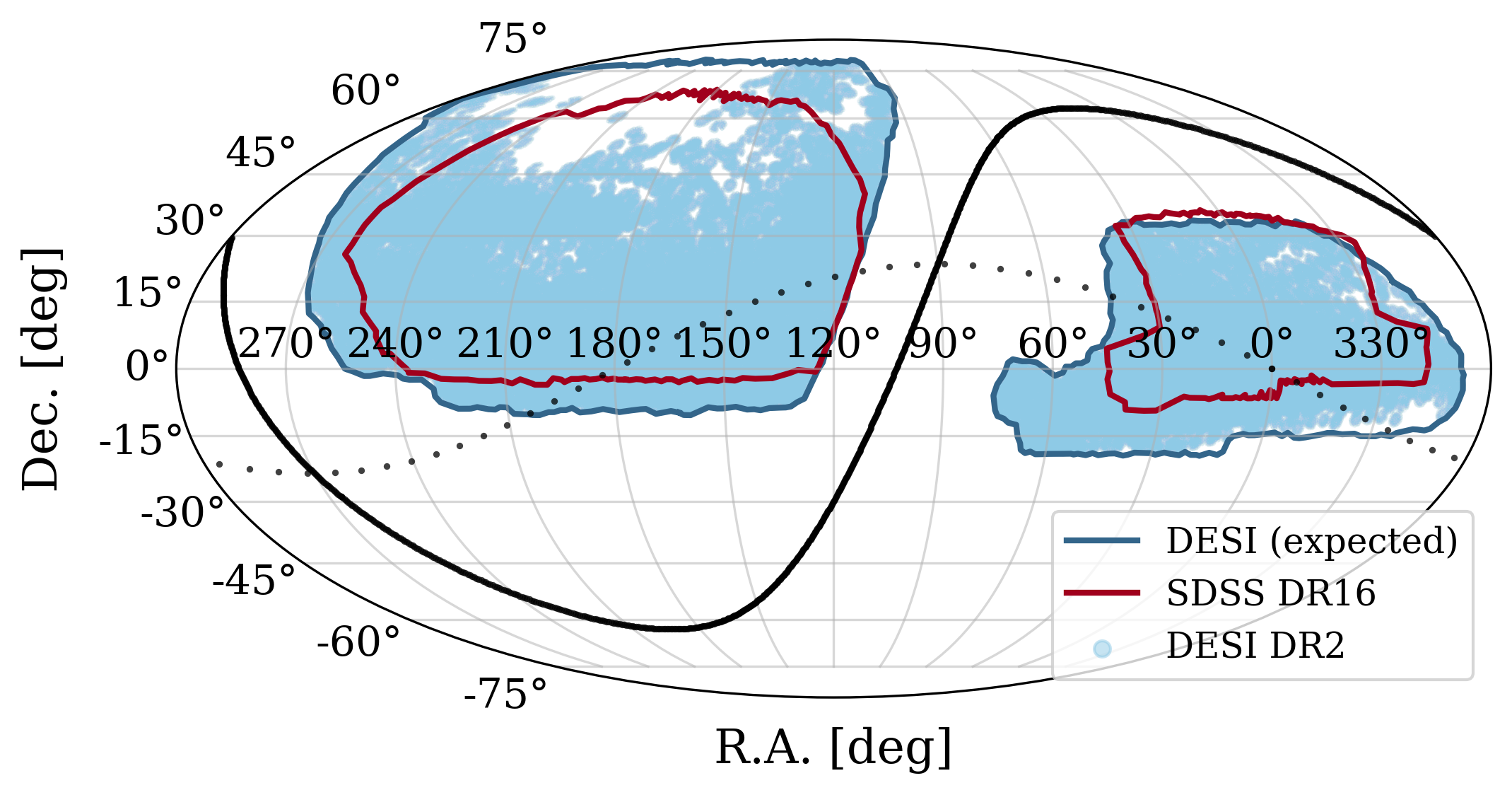}
\end{minipage}
\begin{minipage}{0.29\linewidth}
    \centering
\includegraphics[width=\textwidth]{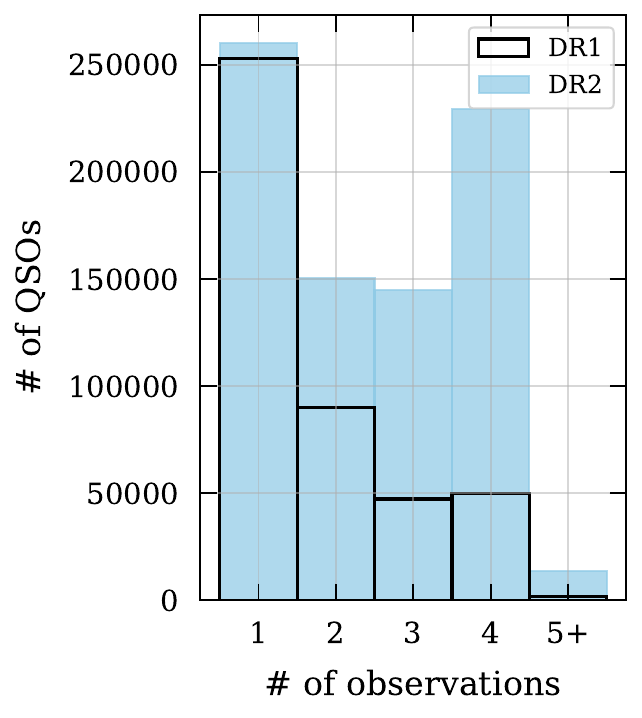}
\end{minipage}
\caption{Left: Expected final DESI (dark blue outline) and the SDSS-DR16 footprint (red outline) together with the spatial distribution of DESI DR2 observed quasars (cyan). For reference we also show the Galactic plane (solid black line) and the Ecliptic plane (dotted black) line. Right: Number of quasars with different numbers of observations for the \lya quasar sample in DESI DR2 (filled cyan) and DR1 (black line).}
\label{fig:desi_footprint}
\end{figure*}

The present paper describes results from the first three years of the survey, which will be part of the future data release 2 (DR2). This dataset represents approximately 70\% of the complete dark time survey.
In addition to the larger area, the typical quasar in DR2 has 2--3 observations, so the spectral SNR per quasar is higher than in DR1. \cref{fig:desi_footprint} shows the DR2 footprint and SDSS DR16 footprint for comparison.

The collaboration plans to release two versions of the data as part of DR2. The \texttt{kibo} data release that was used for most of the analysis validation, and the \texttt{loa} data release, which supercedes \texttt{kibo}, that was used for the final analysis. The \texttt{loa} release has fixes for software bugs related to weighting in coadded spectra, cosmic ray masking, and bookkeeping of quasar redshifts between $1.2 < z < 1.8$. The main impact is that the redshifts of about $\sim 0.5$\% of the quasars with $1.2 < z < 1.8$ changed by more than $100\,\mathrm{km s^{-1}}$ between the two releases.  

\subsection{Spectroscopy} \label{sec:spectra}

\begin{figure*}
    \centering
    \includegraphics[width=\textwidth]{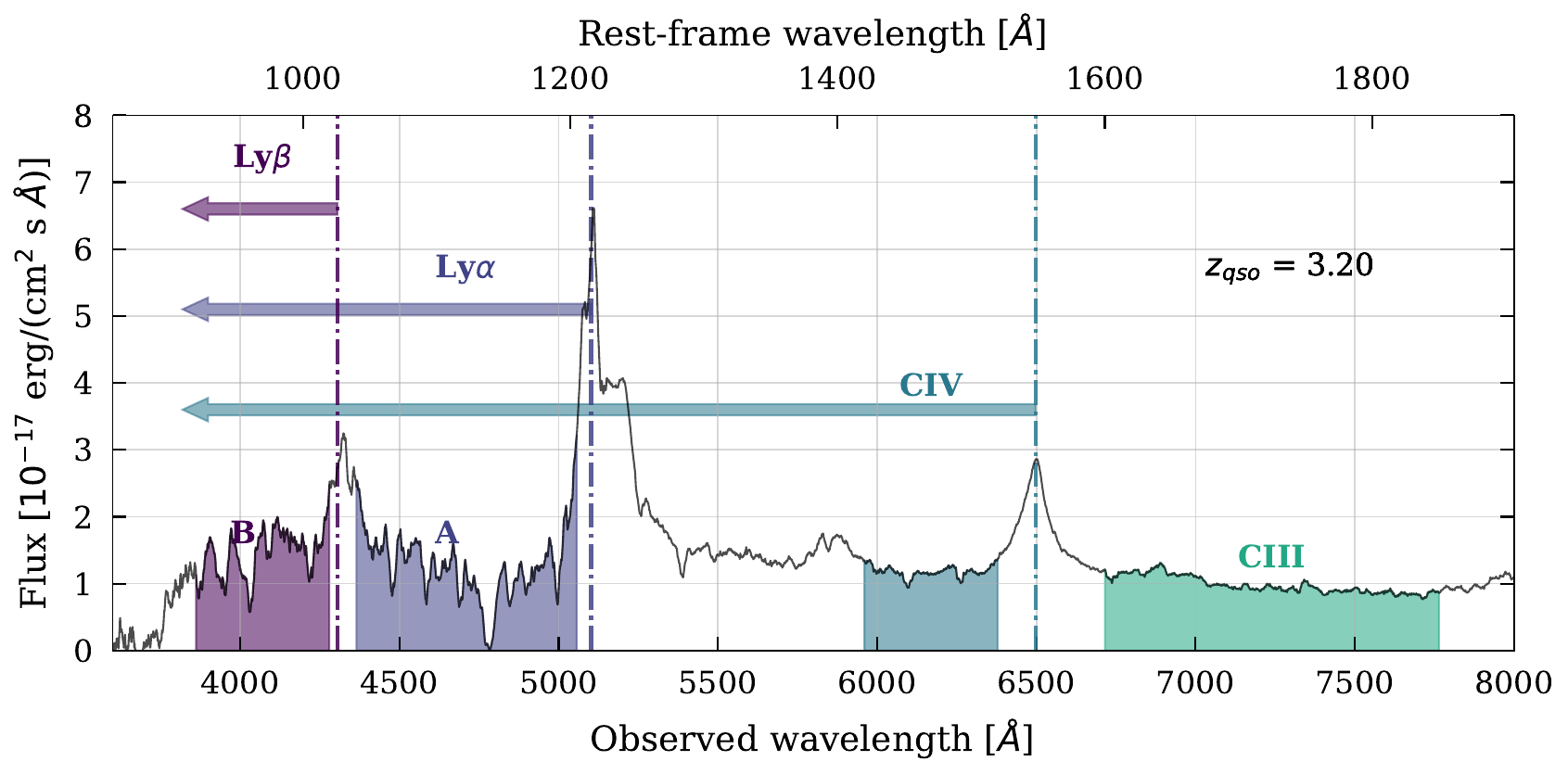}
    \caption{
    Quasar at $z = 3.20$ from the DESI DR2 dataset ($\mathrm{TargetID} = 39627696665266273$). The spectrum shows both regions where we measure the \lya forest: Region A (indigo) extends from 1040 -- 1205 \AA\ in the quasar rest frame and region B (purple) extends from 920 -- 1020 \AA. We measure \lya in both regions. The C\,{\sc IV}  and C\,{\sc III} regions are highlighted in various shades of green. While there is almost no C\,{\sc III} absorption, the C\,{\sc IV}  absorption spans leftward of the C\,{\sc IV}  doublet and thus is a contaminant in both region A and B of the \lya forest. This is an atypically high SNR spectrum and we have smoothed it to better illustrate the quasar spectrum and the forest absorption. This quasar contains a DLA with $\mathrm{log_{10}\,N_{HI}} = 20.55$ at $z=2.93$. }
    \label{fig:desi_spectrum}
\end{figure*}

DESI collects spectra with ten identical spectrographs that record their light from $3600-9800$\,\AA\ in three channels that we refer to as blue, red, and near-infrared. Each of these channels has a separate diffraction grating, and thus each has a distinct range of resolution. The blue channel is the most relevant for the \lya forest and the spectral resolution ranges from approximately $2000-3000$, while the resolution of the red and near-infrared channels ranges from $3500-4500$ and $4000-5500$, respectively. \cref{fig:desi_spectrum} shows an example quasar spectrum from DR2. 

All of the data collected at the observatory are transferred to the National Energy Research Scientific Computing Center (NERSC) for processing and subsequent analysis by collaboration members. The spectra are processed through a spectroscopic pipeline \cite{Spectro.Pipeline.Guy.2023}. This pipeline extracts the spectrum of each object from the two-dimensional data, characterizes the noise, subtracts night sky emission, calculates the wavelength and flux calibration, and ultimately produces one-dimensional spectra with a dispersion of 0.8\,\AA\ per pixel that include information about masked pixels, noise, sky lines, and the spectrograph resolution. All spectra are analyzed with the \texttt{Redrock} software that fits spectral templates and measures redshifts \citep{Redrock.Bailey.2024,Anand2024}. \texttt{Redrock} includes templates for high-redshift quasars \cite{RedrockQSO.Brodzeller.2023}. We also analyze the quasar targets with two additional tools: a Mg\,{\sc II} afterburner \cite{QSO.TS.Chaussidon.2023} that searches for broad Mg\,{\sc II}  emission in any quasar target that \texttt{Redrock} classifies as a galaxy, and \texttt{QuasarNet}, a convolutional neural network quasar classifier originally developed by \cite{Busca18} and optimized for DESI spectra \cite{Green2025.QN}. These two tools identify about 10\% more quasars from the quasar targets relative to \texttt{Redrock}, based on visual inspection during the survey validation period  \cite{VIQSO.Alexander.2023}. 

DESI prioritizes quasars for the study of the \lya forest for up to four observations, and the right panel of \cref{fig:desi_footprint} shows the distribution of the number of observations of the quasars in DR2. In addition, some quasars are observed on multiple nights because a given observation did not achieve the SNR requirement of a single observation (usually due to worsening conditions). When we have multiple observations of a given quasar, we coadd all of the observations to produce the highest SNR spectrum and analyze the coadded spectrum. The coadded spectra are organized based upon their \texttt{HEALPix} pixel \citep{Gorski2005} for our subsequent analysis. 

\subsection{Quasar Catalog} \label{sec:qsocat}

We construct the DESI DR2 quasar catalog in the same manner as for the DR1 catalog described in \DESIIV. Some of the key properties of the catalog are that it contains all of the quasars identified by the three classifiers that have no issues based on the spectroscopic pipeline, with the exception of a low $\Delta \chi^2$ flag that is not relevant for quasars. This corresponds to retaining quasars with only \texttt{ZWARN=0} and \texttt{ZWARN=4}. The completeness and purity should similarly surpass 95\% and 98\%, respectively \citep{RedrockQSO.Brodzeller.2023}, the statistical precision of the redshift measurements should be greater than $\mathrm{150\,km\,s^{-1}}$, and the catastrophic failure rate should be about 2.5\% (4\%) for redshift errors greater than 3000\,(1000)\,$\mathrm{km\,s^{-1}}$. 
The DR2 catalog has 1,289,874 quasars with $z>1.77$ (which could contribute to the \lya\ quasar cross-correlation) and 824,989 with $z>2.09$ (with a \lya\ forest that could contribute to the correlations given our selection criteria). 

\subsection{Ancillary Catalogs} \label{sec:anccats}

We assemble two ancillary catalogs before we measure the \lya forest. Damped \lya absorption (DLA) systems pose a particular challenge to \lya forest analysis and we identify, catalog, and mask DLAs to mitigate their impact. DLAs are important because these systems with neutral hydrogen column densities $N_{HI} > 2 \times 10^{20} \mathrm{cm^{-2}}$ have damping wings that can extend for thousands of $\mathrm{km \, s^{-1}}$ and thus impact the continuum level\footnote{For reference, $1000\,\mathrm{km \, s^{-1}} \approx 7.5 h^{-1}\,\mathrm{Mpc}$ at $z = 2.5$.}.  They are also more strongly clustered than the \lya forest \cite{FontRibera2012b}, which would complicate the model of the correlations. DLAs can consequently compromise a significant fraction  of the \lya forest for distinct objects, and do so with a distinct clustering signature \cite{FontRibera2012a}. For DR2, we developed a new template-based approach to identify DLAs and measure their column density and redshift \cite{Y3.lya-s2.Brodzeller.2025}. That paper also presents the results from running previously-developed codes based on CNN \citep{Wang2022} and Gaussian Process \cite{Ho2021} methods, and describes how we combined measurements from all three algorithms to construct our DLA catalog. We refer to the supporting paper on DLAs \cite{Y3.lya-s2.Brodzeller.2025} for the details of how we constructed the DLA catalog for the DR2 analysis and to the supporting paper on mocks \cite{Y3.lya-s1.Casas.2025} for a study that shows how several alternative, reasonable choices of catalog combinations have a negligible impact on our measurements of the BAO parameters. 

Broad absorption line (BAL) quasars have absorption troughs associated with many emission features, including a number that are within the range of the \lya forest analysis \cite{Ennesser2022}. These features add absorption that is unrelated to the matter distribution in the intergalactic medium. In addition, the absorption of bright quasar emission lines produces redshift errors \citep{Garcia2023,Filbert2023}. We identify BALs in 20.5\% of quasars in the redshift range $1.77 < z < 3.8$ that is used for \lya forest analysis and 17.1\% over the redshift range $1.57 < z < 5$ where we can measure C\,{\sc IV} BALs in DESI data. These percentages are slightly higher than in DR1 (19.8\%, 15.7\%), which we attribute to the higher average SNR of the DR2 quasar spectra. As for DR1, we use the C\,{\sc IV} absorption trough locations to identify and mask the expected locations of BAL troughs associated with other emission lines that fall within the \lya forest. One of the DESI DR1 supporting papers \cite{KP6s9-Martini} describes the method in detail as well as demonstrate the minimal impact of incompleteness in the BAL catalog on the BAO measurements. 

\subsection{Forest Measurements} \label{sec:deltas}

We measure flux decrements in the wavelength range 3600 - 5772\,\AA\ in the observed frame to study the fluctuations due to the \lya forest. The lower bound is set by the minimum design wavelength of the spectrographs and the upper bound corresponds to the middle of the transition region between the blue and red channels that is set by the red dichroic. We measure the forest in two different regions of the quasar rest-frame spectrum, the \lya\ absorption in Region B or \lya(B): 920 - 1020\,\AA, and the \lya\ absorption in Region A or \lya(A): 1040 - 1205\,\AA. Measurements in the B region are generally lower SNR than those in the A region as the B region is only visible in the highest redshift quasar spectra, which are on average fainter, and this region also contains absorption from higher-order Lyman series lines. We therefore compute separate correlations with the A and B regions. Both of these regions are shown on an illustrative quasar spectrum in \cref{fig:desi_spectrum}. That figure also has a label adjacent to the C\,{\sc III} emission line of the region from 1600 - 1850\,\AA\ that we use to quantify spectro-photometric calibration errors. 

We produce \lya forest measurements with the same method described in \DESIIV. Very briefly, we start with the application of four masks that remove bad pixels and astrophysical contaminants. These include cosmic rays, the expected locations of BALs, and the cores of DLAs. We then discard forests that are less than 120\,\AA\ wide in order to have a sufficient path length to fit a model to the continuum. Lastly, we calculate the mean transmitted flux fraction in the C\,{\sc III} region in order to derive a small correction to the spectro-photometric calibration (peaking at $5\%$ at 3650\AA) and instrumental noise estimates (with a 1.5\% correction to the noise variance). 

All of our subsequent analysis uses the transmitted flux field $\delta_q(\lambda)$, where $q$ is index for a given quasar. This is the ratio of the observed flux to the expected flux minus one: 
\begin{equation}
    \delta_{q}\left(\lambda\right) = \frac{f_{q}\left(\lambda\right)}{\overline{F}\left(\lambda\right)C_{q}\left(\lambda\right)} - 1~,
    \label{eqn:delta_definition}
\end{equation}
where $\overline{F}\left(\lambda\right)$ is the mean transmission, $C_{q}$ is the unabsorbed quasar continuum, and $q$ is a given quasar. The procedure to estimate the unabsorbed continuum is described in detail in \cite{Ramirez2024}. This assumes that the product $\overline{F} C_{q}\left(\lambda\right)$ for a given quasar is a universal function of the rest-frame wavelength $\overline{C}\left(\lambda_{\rm rf}\right)$, although corrected by a first degree polynomial in $\Lambda\equiv\log\lambda$ 
to account for quasar luminosity and spectral diversity and the redshift evolution of $\bar{F}(z)$. This product is: 
\begin{equation}
    \overline{F}\left(\lambda\right) C_{q}\left(\lambda\right) = \overline{C}\left(\lambda_{\rm rf}\right) \left(a_{q} + b_{q}\frac{\Lambda - \Lambda_{\rm min}}{\Lambda_{\rm max} - \Lambda_{\rm min}}\right) ~.
    \label{eqn:aqbq_definition}
\end{equation}
 We solve for $\overline{C} (\lambda_{\mathrm{rf}}), a_q,$ and $b_q$ with the maximum likelihood method. This fit also takes into account the pipeline noise, which is corrected to account for small calibration errors, and the intrinsic variance in the \lya forest. 

\section{Correlations} \label{sec:correlations}

\begin{figure*}[ht]
\centering
\includegraphics[width=0.9\textwidth]{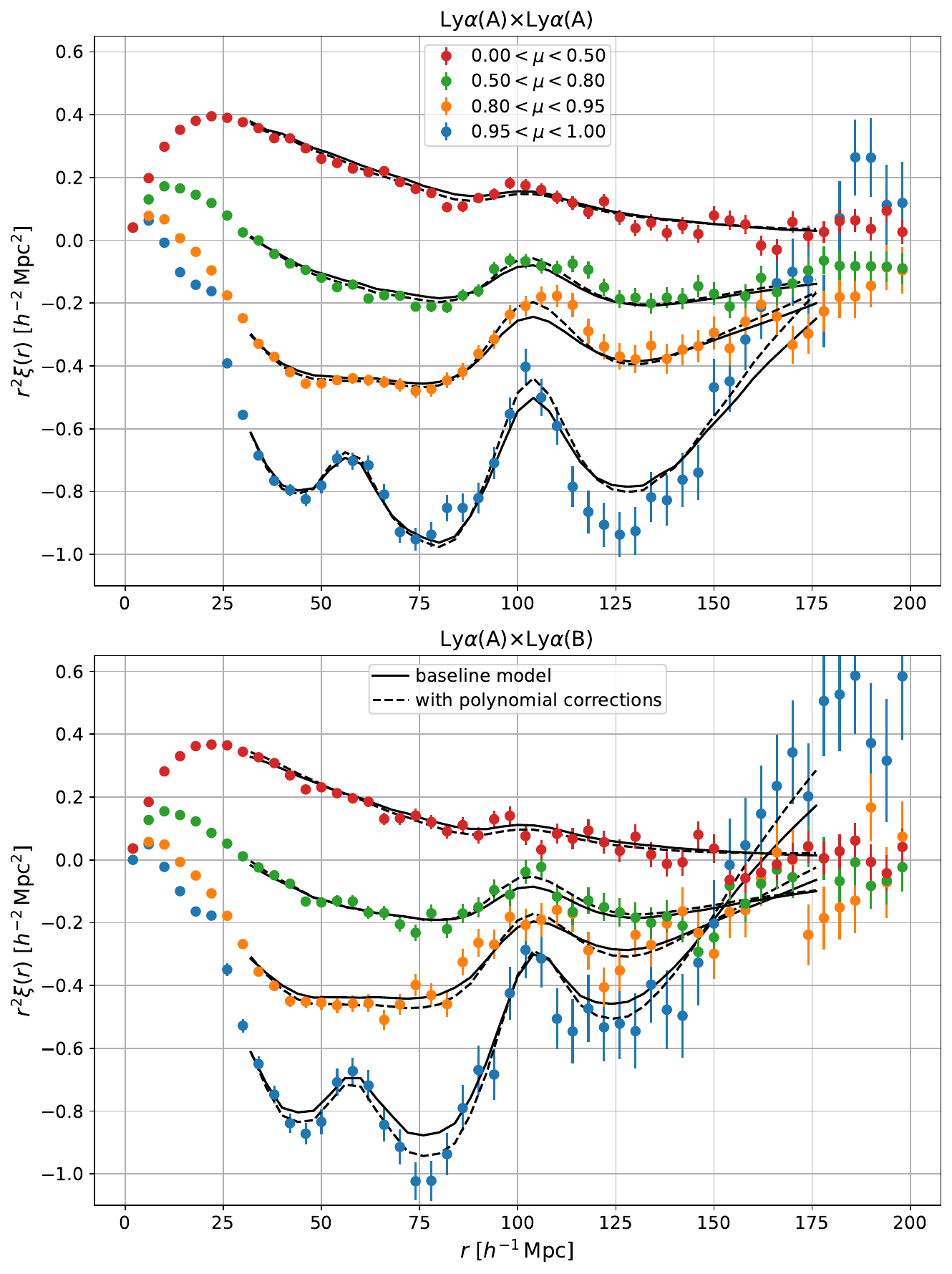}
  \caption{Measured \lyaxlyaA  and \lyaxlyaB auto-correlation functions (top and bottom panel). The different colors and markers correspond to different orientations with respect to the line-of-sight, with blue correlations being close to the line-of-sight $0.95<\mu<1$. The best fit model to all four correlations (see \cref{sec:fit}) is represented with solid curves. The dashed curves show the best fit model with additive broad-band corrections  (see \cref{sec:dataval}). 
  \label{fig:baseline-correlation-lyalya-wedges}}
\end{figure*}

\begin{figure*}[ht]
\centering
\includegraphics[width=0.9\textwidth]{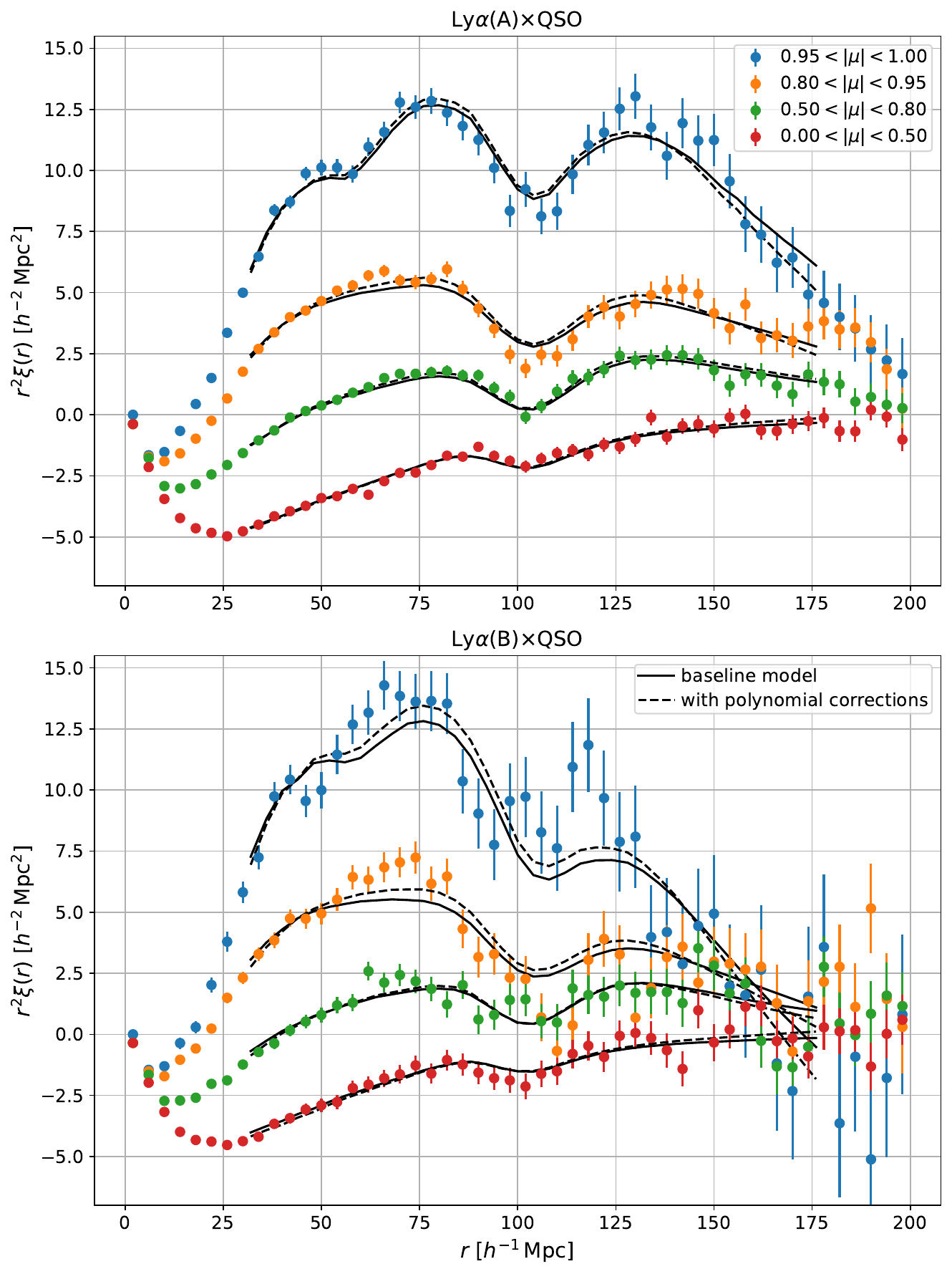}
  \caption{Measured \lyaxqsoA  and \lyaxqsoB cross-correlation functions (top and bottom panel). The different colors and markers correspond to different orientations with respect to the line-of-sight, with blue correlations being close to the line-of-sight $0.95<|\mu|<1$. 
  The best fit model to all four correlations (see \cref{sec:fit}) is represented with solid curves.  The dashed curves show the best fit model with additive broad-band corrections (see \cref{sec:dataval}). 
  \label{fig:baseline-correlation-lyaqso-wedges}}
\end{figure*}

We measure four correlation functions using the quasar catalog and the $\delta_q$ field described in the previous section: the auto-correlation of the \lya forest from region A, hereafter \lyaxlyaA, the auto-correlation of \lya absorption between regions A and B, \lyaxlyaB, and the cross-correlations of \lya from regions A and B with quasars, \lyaxqsoA and \lyaxqsoB. We employ a rectangular grid of comoving separation along and across the line of sight ($r_\parallel$ and $r_\perp$) and use the fiducial cosmology defined in \cref{tab:fid_cosmo} to convert angles and redshifts to comoving separations. The bin size is $4\,\hMpc$, and the measurements extend to separations of $200\,\hMpc$. The data vector size is 2,500 ($50 \times 50$ bins) for \lyaxlyaA and \lyaxlyaB, and twice as large for \lyaxqsoA and \lyaxqsoB because we consider both negative and positive longitudinal separations, where a negative separation is where the quasar is behind the forest pixel. The full data vector size is 15,000 elements.

The measured correlations are shown in \cref{fig:baseline-correlation-lyalya-wedges} and \cref{fig:baseline-correlation-lyaqso-wedges}. We compute these correlations with the same method we used for the DESI DR1 analysis presented in \DESIIV, building on earlier work from BOSS and eBOSS \citep{Bautista2017,dMdB2020}. As in \DESIIV, we present the data in the form of wedges, which means we show the mean correlation in bins of separation $r = (r_\parallel^2 + r_\perp^2)^{1/2}$ and in intervals of the cosine between the line of sight and the separation vector $\mu = r_\parallel / r$. We emphasize that we fit the 2D data (described in \cref{sec:measurement}) and these wedges are just shown for illustrative purposes. \DESIIV (and references therein) contains a more complete description of the correlation function estimators and the measurement of the full covariance matrix. The only changes to the analysis for DR2 are improvements in the computation of the distortion matrix and a minor change to the computation of the metal matrices. We describe these changes in the following subsections.

\begin{table}
\centering
\begin{tabular}{cc}
\hline
\hline
Parameter                              & Planck (2018) cosmology             \\
                                       & (TT,TE,EE+lowE+lensing)                \\
\hline
$\Omega_{\rm m}h^2$ =                  & 0.14297                             \\
$+\Omega_{\rm c}h^2$                   & 0.12                                \\
$+\Omega_{\rm b}h^2$                   & 0.02237                             \\
$+\Omega_{\rm \nu}h^2$                 & 0.0006                              \\
$h$                                    & 0.6736                              \\
$n_{\rm s}$                            & 0.9649                              \\
$10^9 A_{\rm s}$                       & 2.100                               \\

\hline
$\Omega_{\rm m}$                       & 0.31509                            \\
$\Omega_{\rm r}$                       & 7.9638e-05                         \\
$\sigma_8 (z=0)$                     & 0.8119                               \\
$r_{\rm d} \; [\rm Mpc]$             & 147.09                              \\
$r_{\rm d} \; [h^{-1} \rm Mpc]$      & 99.08                               \\
$D_{\rm H}(\zeff=2.33)/r_{\rm d}$   & 8.6172                               \\
$D_{\rm M}(\zeff=2.33)/r_{\rm d}$   & 39.1879                              \\
$f(\zeff=2.33)$                     & 0.9703                               \\
\hline
\end{tabular}
    \caption{
      Parameters of the fiducial flat-$\Lambda$CDM cosmological model used in the analysis to transform observed angular separations and redshifts into physical separations and for modelling, as well as calculate the broadband and peak components of the power spectrum used for the fits. This is the same fiducial model as in \DESIIV. The first part of the table gives the cosmological parameters and the second part gives derived quantities used in this paper.
        }
    \label{tab:fid_cosmo}
\end{table}

\subsection{Distortion Matrix} \label{sec:dm}

The continuum fitting procedure introduces a distortion of the measured \lya fluctuation field \citep{Slosar2011}. This is because it effectively subtracts the mean $\delta$ and the first moment of each forest. The distortion is equivalent to replacing the $\delta_k$ with $\tilde{\delta_i} = \eta_{ik} \delta_k$, where the indices $i$ and $k$ are wavelength indices for the \lya fluctuations along the line of sight of the same quasar, and the coefficients $\eta_{ik}$ are given in Eq.~3.3 of \DESIIV. We perform the same operation on the model correlation function, which is done with the distortion matrix defined below. The  correlation function $\tilde{\xi}$ of the $\tilde{\delta_i}$ in a bin $M$ is the following weighted sum:
\begin{eqnarray}
  \tilde \xi_M &=& W_M^{-1} \sum_{i,j \in M} w_i w_j \tilde \delta_i  \tilde \delta_j \nonumber \\
  &=&  W_M^{-1} \sum_{i,j \in M} w_i w_j \sum_N \sum_{k,p \in N} \eta_{i,k} \eta_{j,p} \delta_k \delta_p \nonumber \label{eq:distort1}
%  &=&  D_{M N} \xi_N \label{eq:distort1}
\end{eqnarray}
Here $W_M = \sum_{i,j \in M} w_i w_j$, and we have omitted the indices of the two quasar lines of sight for the indices $(i,k)$ and $(j,p)$. We have also introduced another set of separation bins $N$, such that we can replace the products $\delta_k \delta_p$ for $k,p \in N$ by their expectation value, which is modeled as the average value of the undistorted correlation function $\xi$ in the separation bin $N$.  
We thereby obtain a linear relation between the distorted and undistorted correlation function and the coefficients of this linear relation are the distortion matrix elements \citep{Bautista2017}.
Note that the comoving separation bin size of the model is $2\,\hMpc$, which is a factor of two smaller than we use for the data ($4\,\hMpc$, as in \DESIIV) and we extend the modeling to $300\,\hMpc$ along $r_\|$.

We improve the modeling of the distortion in this paper by accounting for the redshift evolution of the clustering. We approximate the evolution of the clustering amplitude with redshift with a power-law of $(1+z)$, so the expectation value of the \lya auto-correlation is:
\begin{equation}
  %\left< \delta_k \delta_p \right>_{k,p \in N} = \left( \frac{1+z_k}{1+z_{ref}}\right)^{\gamma_\alpha-1} \left( \frac{1+z_p}{1+z_{ref}}\right)^{\gamma_\alpha-1} \xi_{N}(z_{ref}) \label{eq:distort2}
\left< \delta_k \delta_p \right>_{k,p \in N} = \left( \frac{(1+z_k)(1+z_p)}{(1+z_{ref})^2}\right)^{\gamma_\alpha-1} \xi_{N}(z_{ref}) \label{eq:distort2}
\end{equation}
where  $z_{ref}$ is a reference redshift and $\gamma_\alpha$ is the \lya bias evolution index (the additional term of $-1$ on the power law index in \cref{eq:distort2} accounts for the growth of structure in the matter dominated era). 
The result is a new distortion matrix $D$ that relates the measured correlation function to the undistorted one at $z=z_{ref}$. The elements are:
\begin{eqnarray}
  D_{M N} &=& W_M^{-1} \sum_{i,j \in M} w_i w_j \nonumber \\
  && \sum_{k,p \in N} \eta_{i,k} \eta_{j,p} \left( \frac{1+z_k}{1+z_{ref}}\right)^{\gamma_\alpha-1} \left( \frac{1+z_p}{1+z_{ref}}\right)^{\gamma_\alpha-1}
\end{eqnarray}

The elements of the distortion matrix for the cross-correlation of \lya with quasars are similarly:
\begin{eqnarray}
  D^X_{M N} &=& \left( W^X_M \right)^{-1} \sum_{i,Q \in M} w_i w_Q \nonumber \\
  && \sum_{k,Q \in N} \eta_{i,k} \left( \frac{1+z_k}{1+z_{ref}}\right)^{\gamma_\alpha-1} \left( \frac{1+z_Q}{1+z_{ref}}\right)^{\gamma_{QSO}-1}
\end{eqnarray}
where $W^X_M = \sum_{i,Q \in M} w_i w_Q$ and $\gamma_{QSO}$ is the quasar bias evolution index. 

Our new approach with the redshift evolution in the distortion matrix improves the fit to synthetic data sets (see \cref{sec:mockval}) and marginally improves the fit to the data with $\Delta \chi^2 = -4$.
% 9304.46 instead of 9308.48, see /global/cfs/cdirs/desicollab/science/lya/y3/loa/plots/README
In \cref{sec:dataval} we show that the improvement has a negligible impact on the best fit BAO parameters. A more detailed analysis of the continuum fit distortion will appear in the near future \cite{BuscaRich2025}. 

\subsection{Modeling Metals} \label{sec:mm}

The absorption we observe at a given wavelength in the \lya forest includes contributions from several absorbers with different atomic transitions located at different redshifts along the line of sight (see \cite{Gordon2023} and \DESIIV for more details). In the correlations of the \lya forest, we primarily detect contamination from Si\,{\sc II} and Si\,{\sc III} lines, along with more minor contamination from other foreground absorbers (principally C\,{\sc IV}). 

There is a true comoving separation vector $\mathbf{r}^t_{i,j}$ at the true transition wavelengths for each pair of atomic transitions and for each measured pair $\delta_i \delta_j$ (with two observed wavelengths and a separation angle), as well as a comoving separation $\mathbf{r}^\alpha_{i,j}$ based on the assumption that both absorption features are from the \lya transition. This difference between true and assumed comoving separation results in spurious correlations \cite[e.g.][]{Bautista2017}. For instance, at the effective redshift of our observations ($\zeff=2.33$) and for our choice of fiducial cosmology (see \cref{tab:fid_cosmo}), the correlation between \lya absorption and four Silicon lines: Si\,{\sc III} at 1207\,\AA\ and Si\,{\sc II} at 1190\,\AA\, 1193\,\AA\, and 1260\,\AA, result in spurious peaks in the measured correlation functions at $|\Delta r_{\parallel}| \sim 21$, $59$, $52$ and $104\,\hMpc$, respectively. These metal lines are prominent in stacked spectra centered on strong absorbers \citep{Pieri2014}.

We estimate the contribution from metals by looping over all possible pairs that contribute to the measured correlation function and compute the relative weight of pairs with $\mathbf{r}^t_{i,j} \in B$ among those with $\mathbf{r}^\alpha_{i,j} \in A$, where $A$ and $B$ are separation bins. Those weights define what we call the metal matrix. We use the elements $M_{AB}$ of this matrix to model the observed correlation function. There are terms of the form $\xi^{meas}_A = \sum_B M_{AB} \xi^{true}_B$ for each pair of transitions.

In \DESIIV, we estimated the metal matrix along $r_\parallel$ only and ignored the small variation in $r_\perp$. For that calculation we used the stack of weights as a function of wavelength. We improve on that approach in this work by also evaluating the variation as a function of  $r_\perp$. Now we loop over pairs of wavelengths using the same stack of weights and we integrate over a suitable range of possible separation angles in order to get a representative weighted sample of pairs of $r_\perp^\alpha$ and $r_\perp^t$, which we use to obtain a better estimate of the metal matrix elements. The end result is two matrices, one that addresses the dependence on $r_\parallel$ and one that addresses the dependence on $r_\perp$. We use both to compute the contribution of metals to the measured correlation function. This modification is more accurate than the one used in \DESIIV, although the improvement has a very small impact on the model. The change in $\chi^2$ is smaller than one for more than 9000 degrees of freedom in the fit of the \DESIIV correlation functions, and the change in the BAO parameters is less than 0.1\%. 

\section{Measurement} \label{sec:measurement}

We measure the BAO scale with the same procedure as described in \DESIIV. A key aspect of this procedure is that we use the \texttt{Vega} package\footnote{\url{https://github.com/andreicuceu/vega}} to model the correlations and for the parameter inference. The model for the correlations includes the large-scale power spectrum of fluctuations in the \lya forest that separates the peak (BAO) and smooth components, redshift evolution, and the distortion matrix due to continuum fitting. One significant improvement is that the distortion matrix now accounts for redshift evolution, as described in \cref{sec:dm}. The model also includes contamination from metal absorption in the IGM, correlated noise due to data processing, high column density (HCD) systems, quasar redshift errors, the proximity effect due to the impact of quasar radiation on the IGM, and small scale correlations such as due to non-linear peculiar velocities. We sample a Gaussian likelihood with the Nested Sampler \texttt{Polychord}\footnote{\url{https://github.com/PolyChord/PolyChordLite}} \cite{PolyChord2015a,PolyChord2015b}. The baseline fit is described next in \cref{sec:fit}. While the quality of the fit is formally very good, there are numerous outliers that are apparent from careful inspection of the projections shown in \cref{fig:baseline-correlation-lyalya-wedges} and \cref{fig:baseline-correlation-lyaqso-wedges}. These outliers are difficult to interpret in isolation because of these are just projections of the 2D fits and there are correlations between the bins. We present a detailed analysis of the significance of outliers in \cref{sec:outlier}. 

\subsection{Baseline Fit} \label{sec:fit}

The four correlations are: \lyaxlyaA, \lyaxlyaB, \lyaxqsoA, \lyaxqsoB (see \cref{sec:correlations}). The two auto-correlation functions each have 2,500 data points and the two cross-correlations each have 5,000 data points, for a total of 15,000 data points. We also have a $15,000 \times 15,000$ covariance matrix that accounts for the small cross-covariance between these four correlation functions. We fit these correlations over the range $30 < r < 180\,\hMpc$, which is a somewhat narrower range than the $10 < r < 180\,\hMpc$ range used in \DESIIV. The reason for the smaller fit range is because the model is not a good fit at smaller scales, which are hard to model well. This reduces the total number of data points in the fit to 9306. 

We fit these data points with the two BAO parameters $\alpha_\|$ and $\alpha_\perp$ and 15 nuisance parameters, or 17 parameters total. The nuisance parameters include the bias and redshift space distortion (RSD) parameters for the \lya forest, bias parameters for the quasars and for five metal lines in the forest, bias and RSD parameters for high column density systems (HCDs), a size scale parameter for HCDs, a potential shift in the cross-correlation function, redshift errors, the quasar transverse proximity effect, and a term that accounts for correlated noise in data processing. We added new, informative priors on several of the nuisance parameters due to the change in the fit range.

\begin{figure}
\centering
\includegraphics[width=0.45\textwidth]{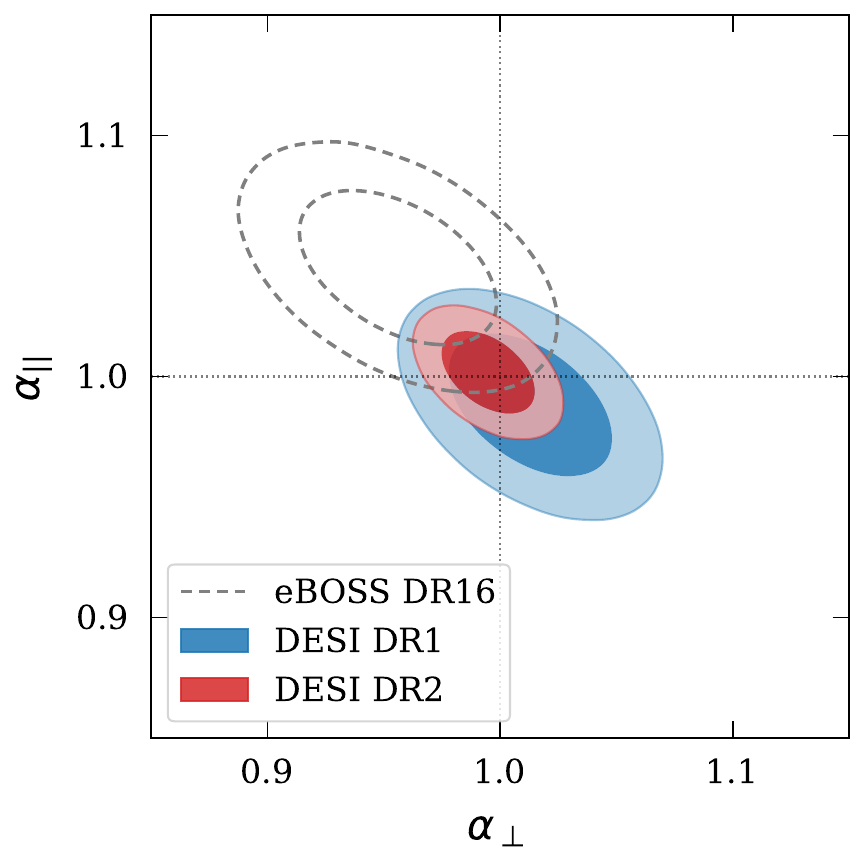}
\caption{\lya BAO measurement of $\alpha_{\|}$ vs.\ $\alpha_{\perp}$ from DESI DR2 (red contour) compared to DESI DR1 (blue contour) and eBOSS DR16 (dashed, gray contours).}
\label{fig:desi_bao_alpha}
\end{figure}

The best fit values of the BAO parameters are:
\begin{equation}
    \alpha_\| = 1.002 \pm 0.011
\end{equation} 
and 
\begin{equation}
    \alpha_\perp = 0.995 \pm 0.013
\end{equation} 
with a correlation coefficient of $\rho = -0.46$. The $\chi^2$ of the best-fit model is $9304.5/(9306-17) = 1.002$, and the probability of having a value larger than this is 45\%. \cref{fig:desi_bao_alpha} shows this BAO measurement of $\alpha_{\|}$ vs.\ $\alpha_{\perp}$ compared to the results from the first DESI data release (DR1) and eBOSS DR16. This figure highlights the two-fold improvement in precision achieved with the new data set. 

We provide further information about the nuisance parameters in \cref{app:nuisance} and the values from our baseline fit are listed there in \cref{tab:baseline}, along with separate measurements from just the two auto-correlations and just the two cross-correlations. ~\cref{sec:dataval} discusses the excellent agreement between the auto- and cross-correlation measurements and the impact of changes to the fitting process on the BAO measurements. 

\subsection{Goodness of fit} \label{sec:outlier}

In spite of the quality of the $\chi^2$ for the baseline fit, the wedges in \cref{fig:baseline-correlation-lyalya-wedges} and \cref{fig:baseline-correlation-lyaqso-wedges} show that the best fit model does not go through all of the points. There is a slight mismatch in the broadband part of the \lyaxlyaA and \lyaxlyaB correlation functions at separations $r$ between 30 and $80\,\hMpc$ and $\mu<0.95$, where the best fit model lies above the data points on average. We suspect that this is caused by an imperfect modeling of the correlations at smaller separations ($<30\,\hMpc$) that affect the larger separations because of the distortion induced by the continuum fitting. We plan to investigate the impact of continuum fitting in upcoming analyses thanks in particular to recent improvements in the continuum prediction from the red side of the quasar spectra \citep{Turner2024} that are not affected by these distortions. For now, we resort to verifying that this imperfect modeling does not bias the BAO measurements by adding polynomial corrections to the model \citep[as in][]{Bautista2017, dMdB2020, DESI2024.IV.KP6}, which are shown with the dashed curves in \cref{fig:baseline-correlation-lyalya-wedges} and \cref{fig:baseline-correlation-lyaqso-wedges}. These corrections add  Legendre polynomials $L_J(\mu)$ with order $j=0,2,4,6$ divided by powers of $r_i$ with $i=0,1,2$ to each correlation function. There are thus 12 additional parameters for each correlation function, or 48 total. This variation provides a visually better fit to the data and the $\chi^2$ is also somewhat better: $\chi^2/(N_{data}-N_{param})=9243.6/(9306-53)=0.999$ with a p-value of $0.53$. There is not a large improvement in the reduced $\chi^2$, although some of the polynomial coefficients do deviate significantly from zero with more than $3\,\sigma$ significance. We show in \cref{sec:dataval} that the polynomial corrections have a negligible effect on the BAO measurement.

In addition to this imperfect modeling of the distorted continuum, there is also an apparent mismatch between the measured and best fit BAO peak position in the \lyaxlyaA correlation function (top panel of \cref{fig:baseline-correlation-lyalya-wedges}) for the wedge $0.80<\mu<0.95$. We interpret this mismatch as a statistical fluctuation because the best fit BAO peak position is constrained by the ensemble of the data and not just this wedge. 
We have verified this interpretation with independent fits of the BAO peak in each of the four wedges of the \lyaxlyaA correlation function shown in \cref{fig:baseline-correlation-lyalya-wedges}. While the best fit BAO peak position is shifted to larger values for the $0.80<\mu<0.95$ wedge, it is compensated by a shift of the best fit BAO peak position toward lower values for the $\mu>0.95$ wedge. We also note that the results from each of the four wedges are statistically consistent with one another, and we have verified that this mismatch cannot be explained by a fitter convergence issue (including with extensive tests of the fitter on synthetic data, see \cref{sec:mockval}). 

\begin{figure}
\includegraphics[width=0.99\columnwidth]{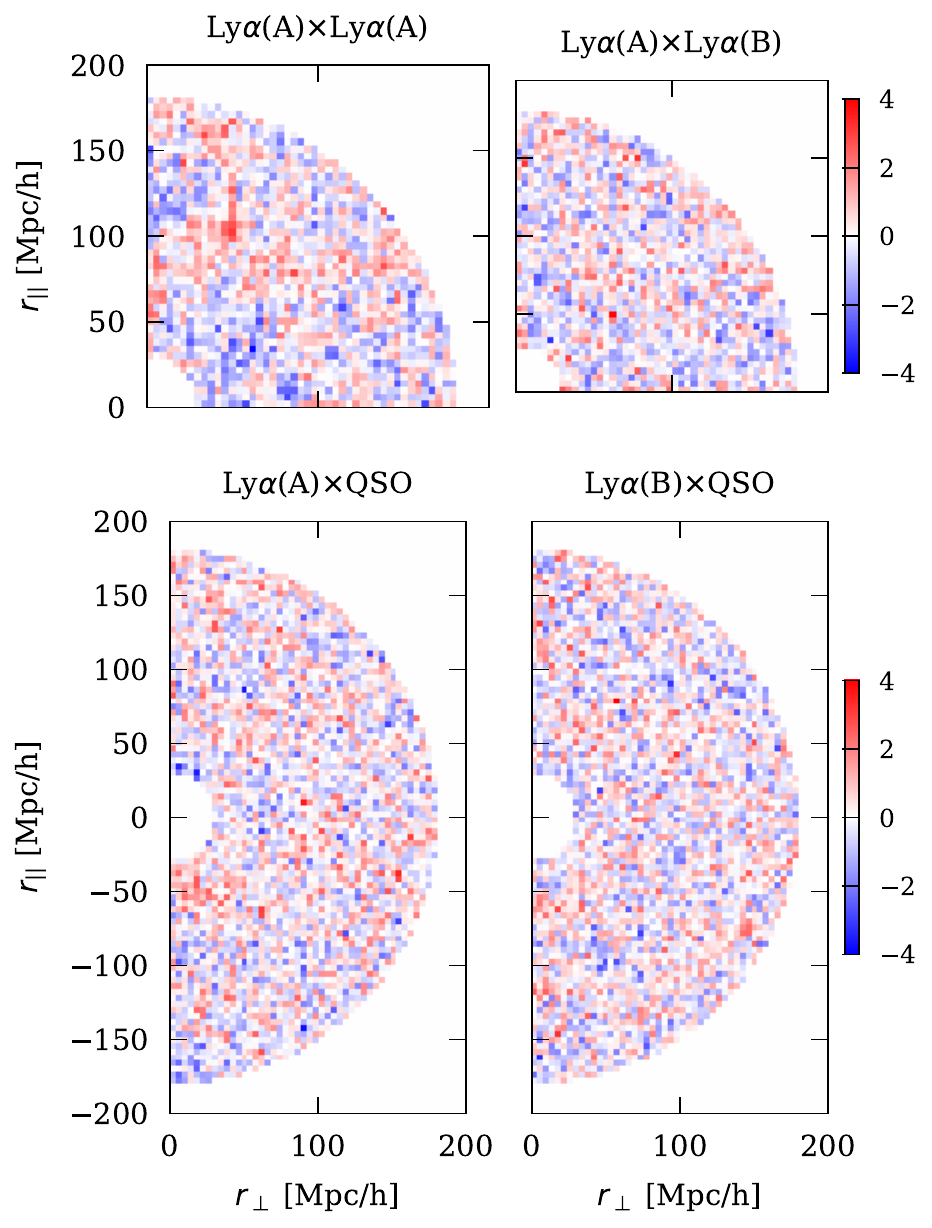}
  \caption{Normalized residuals $(\xi_D - \xi_M)/\sigma_{\xi}$ or the four correlation functions, shown in the 2D space of $r_\parallel$ and $r_\perp$ separation that we use for the fits. 
  \label{fig:residuals2d}}
\end{figure}

We next explore the possibility of an unidentified contamination of the signal with an inspection of the fit residuals and their statistical significance in 2D. \cref{fig:residuals2d} shows the normalized residuals in the 2D plane of comoving coordinates $r_\parallel$ and $r_\perp$ for the four correlation functions: 
$(\xi_D-\xi_M)/\sigma_\xi$, where $\xi_D$ is the measured correlation, $\xi_M$ is the best fit model, and $\sigma_\xi$ is the uncertainty of each element in the correlation function (i.e. square root of the diagonal of the covariance matrix). There is a slight excess correlation in the \lya\ auto-correlation along a line at $r_{\perp} \sim 50 \, \hMpc$, and extended over the range $100  \, \hMpc < r_{\parallel} < 140  \, \hMpc$ where the bin values are individually not significant (2-$\sigma$ excess on average). 
We note that our measurements naturally result in extended correlations along $r_{\parallel}$ due to  imperfections in the continuum fitting. Rejecting the quasar lines of sight with the largest deviations in flux decrement does not change this pattern significantly, so it is not caused by a subset of the data that we could easily isolate. For those reasons, we consider that this structured excess in the residuals is consistent with a statistical fluctuation.

\begin{figure}
\includegraphics[width=0.99\columnwidth]{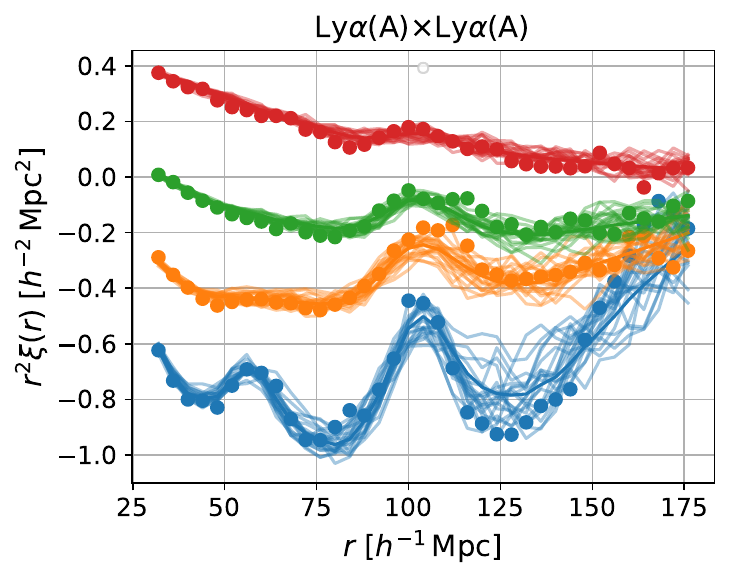}
  \caption{Baseline \lyaxlya\ correlation function (circles) as in \cref{fig:baseline-correlation-lyalya-wedges} along with several curves corresponding to the best fit model plus random noise realizations. The three other correlations are similar. 
  \label{fig:outlier}}
\end{figure}

It is difficult in general to evaluate the agreement between the data and the model from visual inspection of the wedge plots because of the large covariance between neighboring data points: a statistical fluctuation will always look like a bump or a dip extended over several separation bins. In order to help visualize this covariance, \cref{fig:outlier} shows the \lyaxlya wedges with several curves that illustrate the best fit model with additive random realizations of the measurement noise. Those realizations are obtained with a bootstrap technique: each realization is based on a different stack of HEALPix pixels, drawn randomly (with replacement) from the list used to compute the correlation function of the data. A number of these realizations appear to have larger outliers than the data. 

It is also not unlikely to get a large outlier by chance given the large number of points shown in the plots (the famous look-elsewhere effect). Quantitatively, we find worse outliers in 15\% of random realizations of the \lyaxlyaA correlation function wedges compared to the data (and similarly 44\%, 65\% and 11\% worse outliers in random realizations of the \lyaxlyaB, \lyaxqsoA, and \lyaxqsoB  correlation functions, respectively). We conclude that we do not measure a significant disagreement between the data and the best fit model.

\section{Validation} \label{sec:validation}

Most of the methodology for measuring BAO with the \lya forest was carefully developed and validated over many years with applications to many data releases of SDSS and most recently the analysis of the DESI DR1 dataset (\DESIIV). Yet as the \lya forest dataset for DESI DR2 is significantly larger than for DESI DR1, and we have also improved the analysis in a number of ways, we validate the methodology relative to the greater statistical precision we expect from DR2 with synthetic and real data. Before we ran analysis tests on data, we blinded our measurements of the correlation function with the same approach we applied for DR1, although with different shifts. We required that all analysis variations should produce shifts smaller than $\sigma/3$ of the statistical uncertainty (accounting for the correlation between $\alpha_\|$ and $\alpha_\perp$), or that we should understand the reason for any shift (e.g. significant sample size variations). We also ran extensive tests on synthetic datasets with the requirement that the analysis on the mocks produce results that are unbiased by less than this same $\sigma/3$ threshold. Any larger shift that we could not explain would be added to our error budget as a systematic uncertainty. 

The first subsection below describes our tests with synthetic data. Many of these tests were conducted with new synthetic datasets that are described in detail in the supporting paper on mocks \cite{Y3.lya-s1.Casas.2025}. At the time of unblinding in December 2024, we did appear to have a systematic bias of order $\sigma/3$ in the measurement of synthetic data, although after unblinding we traced this systematic bias to the implementation of redshift errors in the synthetic data and were able to decrease the bias to substantially below our threshold. In \cref{sec:dataval} we present a subset of the tests on the blinded data. We did many additional tests that are described in \cref{app:datavalidation}. These repeat most of the tests we performed for DR1 \cite{DESI2024.IV.KP6}.

\subsection{Validation with synthetic data} \label{sec:mockval}

Our validation process with synthetic data largely parallels the extensive work for DR1  \cite{KP6s6-Cuceu}. That \lya BAO measurement was validated with a total of 150 synthetic realizations (or mocks): 100 realizations of the \texttt{LyaCoLoRe} mocks \citep{Farr2020_LyaCoLoRe,Ramirez2022} and 50 realizations of the \texttt{Saclay} mocks \citep{Etourneau2023}. These mocks matched the bias evolution of the forest, the angular, redshift, and magnitude distribution of quasars in DESI DR1, and included different types of astrophysical contaminants, such as metal absorbers, BAL quasars, and DLAs \citep{2024arXiv240100303H}. 

We extended that work to DR2 with improved mock datasets, a larger number of mocks, and more validation tests. These are all described in the companion paper \cite{Y3.lya-s1.Casas.2025}. One significant change is that we have doubled the number of \texttt{Saclay} mocks, from 50 to 100. The second significant change is that we have improved the mocks based on \texttt{LyaCoLoRe} in order to have more realistic quasar clustering on small scales and to simulate the broadening of the BAO feature caused by the non-linear growth of structure. We have generated 300 of these new mocks, and we refer to them as the quasi-linear mocks, or \texttt{CoLoRe-QL} for short.

\begin{figure}
\centering
\includegraphics[width=\columnwidth]{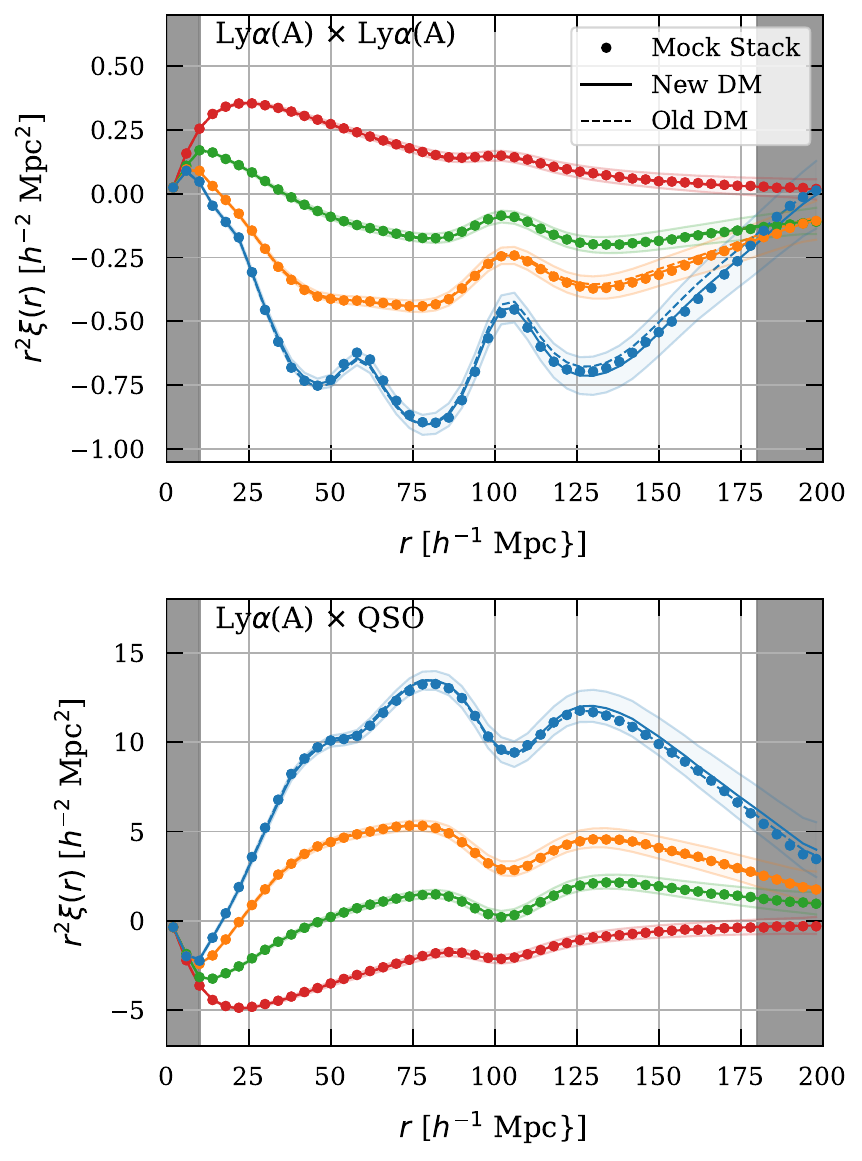}
\caption{Average measurement of the \lya auto-correlation (top) and cross-correlation (bottom) from 300 \texttt{CoLoRe-QL} mocks (dots), compared to the best-fit model using the previous (dashed) and the improved (solid) calculations of the distortion matrix (DM). The shaded areas around the solid line indicate the errorbars from the DESI-DR2 measurements, and the gray shaded areas highlight the scales not included in the fit (the fit to observations starts at $30~ \hMpc$).
} \label{fig:wedges_fit_mocks}
\end{figure}

\cref{fig:wedges_fit_mocks} shows the average measurement (or stack) of the \lya auto-correlation from 300 new \texttt{CoLoRe-QL} mocks, compared to the best-fit model when fitting the correlations from $10~\hMpc$ to $180~\hMpc$ (solid lines). As discussed in the previous sections, in our baseline analysis we fit only separations larger than $30~\hMpc$, but this figure shows that the new method to estimate the distortion matrix (introduced in \cref{sec:dm}) results in a good fit even on smaller scales. 
For comparison, the best-fit model with the previous distortion matrix is shown in dashed lines. 

\begin{figure}
\centering
\includegraphics[width=\columnwidth]{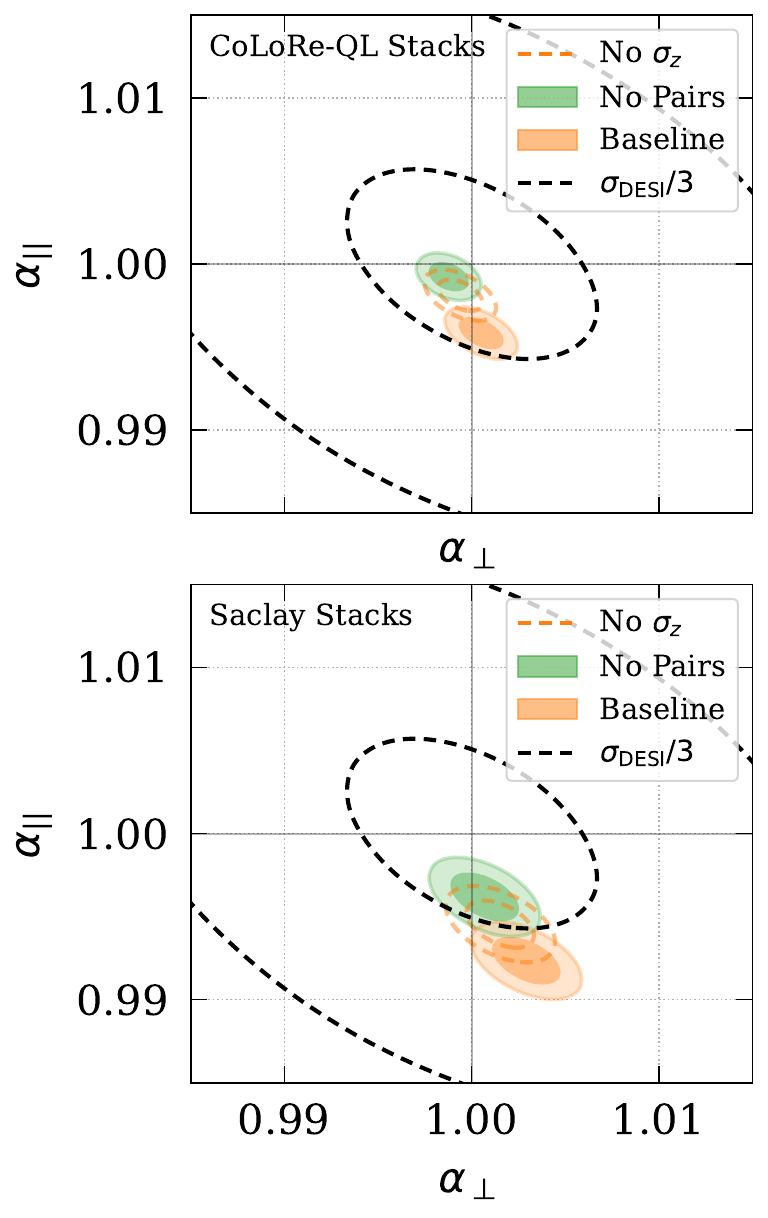}
\caption{BAO fits from the stack of 300 \texttt{CoLoRe-QL} mocks (top) and 100 \texttt{Saclay} mocks (bottom). The baseline configuration (filled green contours) shows a small but significant bias when redshift errors are introduced before continuum fitting. The bias is significantly smaller when redshift errors are added after continuum fitting (empty contours) or when we do an alternative analysis that does not include close pairs (orange contours). The dashed ellipses represent the size of $\sigma/3$ uncertainties based on the DR2 measurement. See \cref{sec:mockval} for further details.}
\label{fig:bao_stack}
\end{figure}

In \cref{fig:bao_stack} we show BAO results from the fit of the four stacked correlation functions from 300 \texttt{CoLoRe-QL} mocks (top), and the stack of correlation functions from 100 \texttt{Saclay} mocks (bottom). In this analysis we have used the same cosmology that was used to generate the mocks, and since these are log-normal mocks (without non-linearities or bulk flows), the true BAO parameters should be $\alpha_\perp=\alpha_\parallel=1$. However, it is clear that our baseline analysis (filled green contours) has a small but significant bias in $\alpha_\parallel$ of order 0.5\%. This bias was already identified in the validation of DESI DR1 with mocks \citep{KP6s6-Cuceu}, although it is more statistically significant now due to the larger statistical power of DESI DR2 and the larger number of mocks in our current study.

As discussed in \cite{Y3.lya-s1.Casas.2025}, the bias is caused by spurious correlations that arise due to quasar redshift errors which produce a smearing effect during the continuum fitting process. As part of this process, we compute a mean quasar continuum, which actually includes numerous, weak, broad emission lines that are present in the forest region. Redshift errors smear these emission lines, and the resulting systematic errors in the mean continuum give rise to spurious correlations. This effect was first discussed in \cite{Youles2022}, and is present in both the \lya\ auto-correlations and \lya-quasar cross-correlations. When redshift errors are only added to the mocks after continuum fitting, the bias is significantly reduced (see the empty contours in \cref{fig:bao_stack}).

In practice, the impact of the spurious correlations is somewhere in between the baseline and ``No $\sigma_z$'' results shown in \cref{fig:bao_stack}. This is because the implementation of redshift errors in the mocks assumes that the Doppler shifts of the emission lines in the forest region are uncorrelated with the Doppler shifts of the broad emission lines at longer wavelengths, which are used to measure the quasar redshifts. Quasar redshifts are well known to have more dispersion than galaxy redshifts, and there are clear systematic errors when redshifts are measured from higher-ionization species like C\,{\sc IV} at longer wavelengths than the quasars \lya emission line. However, if these redshift errors are correlated with the forest region, then the resulting spurious correlations will be overestimated. Therefore, our baseline mock results include the most extreme version of this effect.

Recently, \cite{Gordon2025} introduced a test to gauge the impact of this contamination. This test relies on the fact that for quasar-pixel pairs, the contamination is strongly dependent on the small-scale correlation between the quasar and the host quasar of the forest which contains the pixel \citep[as shown by][]{Youles2022}, while for pixel-pixel pairs it depends on the small-scale cross-correlation between the host quasar of one of the pixels and the other pixel \citep[as shown by][]{Gordon2025}. Therefore, discarding these close pairs would remove most of the spurious correlation, while having a minimal impact on our statistical uncertainty due to the sparsity of quasars. This is confirmed in \cite{Y3.lya-s1.Casas.2025}, which shows that discarding very close pairs significantly reduces the bias in the BAO constraint when the mean continuum is affected by redshift errors. This is also shown with the orange contours in \cref{fig:bao_stack}. We have also performed the same test on the data, and found the impact on BAO is well within our $\sigma/3$ threshold, which demonstrates that the spurious correlations caused by redshift errors do not have a significant impact on our measurement (see \cref{sec:dataval}).

Finally, in \cite{Y3.lya-s1.Casas.2025} we show that the distribution of best-fit BAO values from each individual mock is consistent with the reported uncertainties, and that these uncertainties are similar to the ones obtained in DESI DR2. In particular, the new \texttt{CoLoRe-QL} mocks now produce more realistic BAO uncertainties compared to the mocks used for DESI DR1. The mock uncertainties are now consistent with the uncertainty measured from the data, due to the use of an input power spectrum with a smoother BAO peak, which mimics the non-linear broadening of the peak \citep[see Figure~9 of][]{Y3.lya-s1.Casas.2025}.

\subsection{Validation with blinded data} \label{sec:dataval}

The validation tests with the DESI DR2 data take two forms. The first are ``data splits'', where the dataset is split into two, and the second are ``alternative analyses'', where we explore how various changes in the methodology impact the BAO results. We performed and passed these tests before we unblinded with the internal data release called \texttt{kibo} and then reran the data validation tests after unblinding on a new internal data release called \texttt{loa}. The differences between these two internal data releases are quite minor (see \cref{sec:data}), and the differences between the pre- and post-unblinding tests results are negligible. 

\begin{figure}
\centering
\includegraphics[width=0.75\columnwidth]{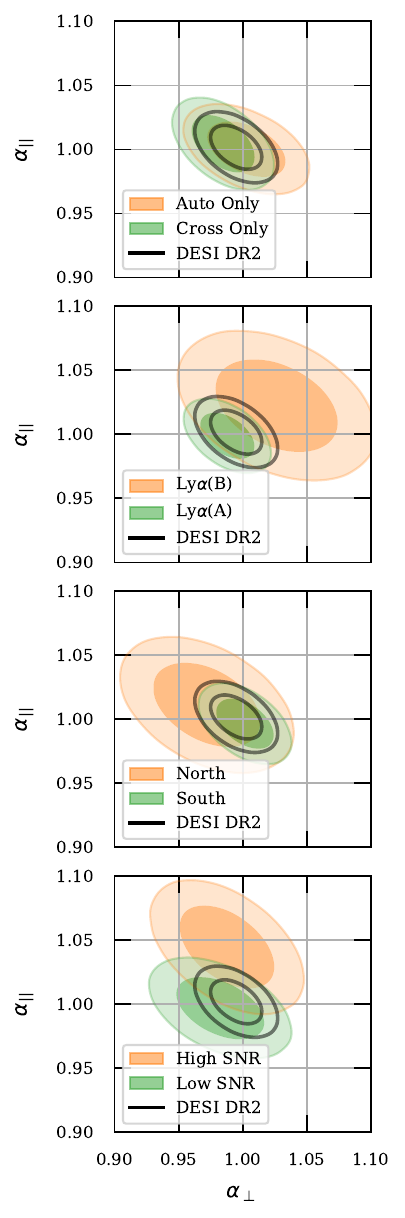}
\caption{Consistency checks of the BAO constraints from four subdivisions of the dataset (orange, green) relative to the DR2 baseline (black, unfilled). From top to bottom the four subdivisions are: 1) Only the auto-correlations (orange) and only the cross-correlations (green); 2) Only Region B (orange) and only Region A (green); 3) Only quasars selected from the northern imaging dataset (orange) and only the southern region (green); 4) Only higher SNR spectra (orange) and only lower SNR spectra (green). These consistency checks are described further in \cref{sec:dataval}. 
}
\label{fig:data_splits}
\end{figure}

\cref{fig:data_splits} shows the robustness of our BAO measurement for four data splits. The most significant test is shown in the top panel. This compares the BAO measurements from just the auto-correlation function with just the cross-correlation function. These results agree with each other very well, and also with the combined result for DESI DR2. The separate BAO measurements from these two alternatives are listed in \cref{tab:baseline}. The second panel also splits the four correlation functions in two with a comparison of the correlations that just include Region A (\lyaxlyaA, \lyaxqsoA) compared to the correlations that just include Region B (\lyaxlyaB, \lyaxqsoB). This data split also demonstrates good consistency, as well as illustrates the much greater statistical power of Region A relative to Region B. 

The two lower panels show tests where we have split the quasar catalog into two, and we have done end-to-end analyses to each subset independently. The third panel separates quasars targeted using the Bok and MzLS surveys (North, declination $\delta > 32.375^\circ$) vs.\ those targeted DECam data (South). The bottom panel shows quasars with high signal-to-noise ratio (SNR $>4.25$) vs.\ low SNR\footnote{We define SNR as the mean signal-to-noise over the rest-frame wavelength range $1420 < \lambda_{\rm rest} < 1480$ \AA.}. We chose this SNR split to achieve approximately equivalent uncertainties in the BAO measurement between the two samples, and not equal numbers of \lya forests. Since the SNR criterion only impacts the forest calculation, the quasars for the cross-correlation measurement were randomly assigned to one of the two samples. The BAO scale parameters are statistically consistent for these two data splits as well.

\begin{figure}
\centering
\includegraphics[width=\columnwidth]{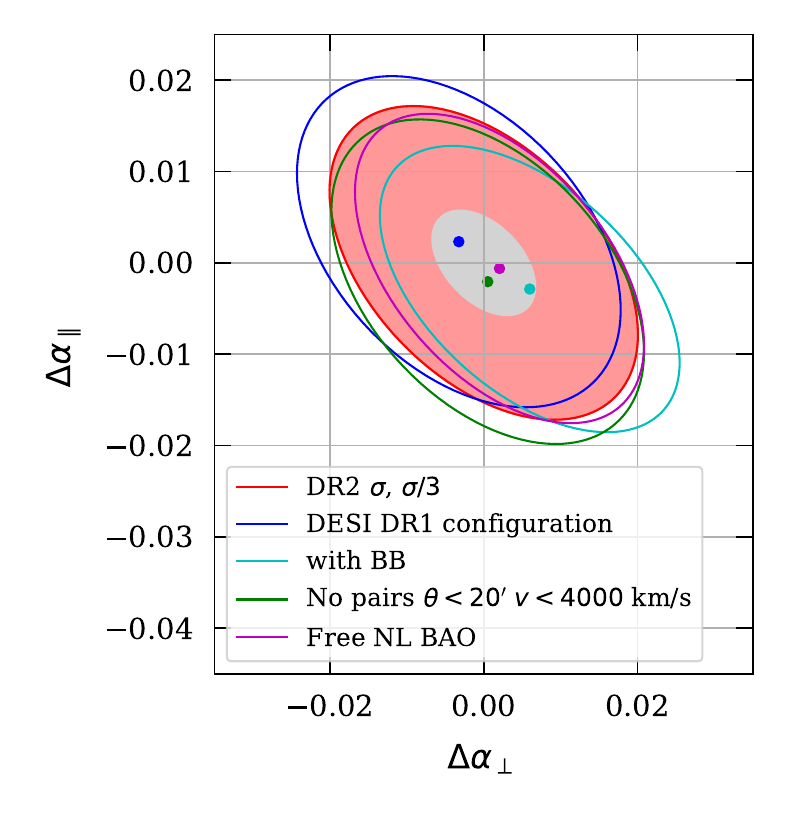}
\caption{Impact of several analysis variations on the $\alpha_\|, \alpha_\perp$ contours relative to the DR2 baseline. The filled contours show the size of the $1\sigma$ statistical errors from \cref{fig:desi_bao_alpha} (red) and the $\sigma/3$ criterion we adopted for further investigation of analysis variations. The variations shown are: analysis with the DR1 configuration (blue), with the addition of broad-band polynomials (cyan), removal of pairs as in the mocks (dark green), and when we rerun the DR2 baseline analysis to also solve for non-linear broadening (magenta).
}
\label{fig:var_2d}
\end{figure}

We have recomputed the BAO measurements with a large number of alternative analysis choices, and the vast majority were previously performed in \DESIIV for DR1. Our standard to compute an analysis alternative was that it be a reasonable alternative to our baseline choice. \cref{fig:var_2d} shows the change in $\alpha_\|$, $\alpha_\perp$ for four of these variations, along with the $1\sigma$ (red) and $\sigma/3$ (gray) contours from the baseline DR2 analysis for reference. The blue ellipse and filled circle shows the change when we rerun with the DR1 analysis configuration. This is a key test that demonstrates that the analysis changes described in \cref{sec:measurement} do not change the BAO measurements, although they do improve the fit to the correlation function. The DR1 configuration includes scales as small as $r_{min} = 10\,\hMpc$, rather than $r_min = 30\,\hMpc$ for the DR2 baseline. The variation labeled BB represents the BAO measurement if we include a broadband polynomial as part of the fit as in \DESIIV. The fit with the BB is shown as the dashed lines in \cref{fig:baseline-correlation-lyalya-wedges} and \cref{fig:baseline-correlation-lyaqso-wedges}. The third shows the change in the BAO measurement if we eliminate the close pairs discussed in \cref{sec:mockval} (angular separations less than $\theta < 20$\,arcminutes and velocity differences of $v < 4000\,\mathrm{km\,s^{-1}}$). Elimination of these close quasar--pixel pairs substantially mitigates a bias in the mocks that is discussed in \cref{sec:mockval} and a supporting paper \cite{Y3.lya-s1.Casas.2025}. The final variation shows the change if we fit for the amount of non-linear broadening of the BAO peak to our fit, instead of fixing it to the prediction from Lagrangian Perturbation Theory as done in \DESIIV. All of these analysis variations produce shifts that are smaller than our threshold of $\sigma/3$ to trigger more thorough analysis. We discuss the remaining data validation tests in \cref{app:datavalidation}. 

\section{Discussion} \label{sec:discussion}

We presented the DR2 baseline measurement of the BAO parameters $\apar$ and $\aper$ in \cref{sec:measurement} at an effective redshift of $\zeff = 2.33$. These parameters correspond to: 
\begin{eqnarray}
\apar &= \frac{D_H(\zeff) / r_d}{\left[ D_H(\zeff) / r_d \right]_{\rm fid}} ~,  \nonumber \\ 
\aper &= \frac{D_M(\zeff) / r_d}{\left[( D_M(\zeff) / r_d \right]_{\rm fid}} ~,
\end{eqnarray}
where $D_M(z)$ is the transverse comoving distance, $D_H(z)=c/H(z)$, and $r_d$ is the sound horizon at the drag epoch. Quantities with the subscript \texttt{fid} are computed with the fiducial cosmology listed in \cref{tab:fid_cosmo}. In this section we discuss the distance measurement from the \lya forest and compare to previous work. We also discuss recent work on the expected size of the shift in the BAO peak due to contributions from nonlinear clustering. 

\subsection{BAO Shift} \label{sec:baoshift}

There are extensive studies of a shift in the BAO position due to non-linear evolution in the galaxy clustering literature. 
These studies find that the shift is typically of order $\lesssim0.5\%$,
and that the shift is greatly reduced after reconstruction (see e.g. \cite{KP4s2-Chen} and references therein).
This BAO shift occurs when the data are fit with a model that does not incorporate the appropriate non-linear corrections.
During the development of our DR2 analysis, the first studies appeared in the literature that reported values for the BAO shift in the \lya\ forest  \citep{Sinigaglia2024,deBelsunce2024,Hadzhiyska2025}. As our analysis provides the first sub-percent \lya\ BAO measurement, and there is no reconstruction applied to the \lya\ field, the impact of a \lya\ BAO shift could be important. In this section we summarize the recent literature on the BAO shift in the \lya forest, provide an estimate of the present theoretical uncertainty in the shift, and describe how we add this as a systematic uncertainty to our analysis. 

Two studies of the shift \cite{Sinigaglia2024,Hadzhiyska2025} used approximate methods based on the Fluctuating Gunn-Peterson Approximation (FGPA) to paint \lya forest skewers onto dark matter fields obtained from either Augmented Lagrangian Perturbation Theory \citep[ALPT,][]{Kitaura:2013} or N-body simulations, respectively. These approaches make it possible to simulate sufficiently large volumes to precisely measure the BAO position, although at the cost of less accurate small-scale clustering. The third study \cite{deBelsunce2024} instead used an effective field theory (EFT) approach \cite{Ivanov:2024} to fit \lya power spectrum measurements from hydrodynamical simulations \cite{Chabanier:2024} and then used the measured EFT parameters to predict the expected BAO shift\footnote{Similar to the approach presented in \cite{KP4s2-Chen} in the context of galaxy clustering}. This has the advantage of relying on hydrodynamical simulations, which provide accurate small-scale clustering. However, these simulations are not large enough to directly obtain a precise BAO constraint \cite{Chabanier:2024}, which is why \cite{deBelsunce2024} rely on the EFT approach.

Both \cite{KP4s2-Chen} and \cite{deBelsunce2024} showed that the BAO shift is sensitive to the relation between the linear and quadratic bias parameters in the EFT approach. Therefore, results based on hydrodynamical simulations should in principle provide more accurate constraints on the BAO shift compared to approximate methods of simulating the \lya\ field. However, the EFT approach \cite{deBelsunce2024} did not study the impact of various simulation modeling choices on the measured bias parameters and the resulting BAO shift. For example, the model for Helium reionization and fluctuations in the UV background are likely to have an impact on these parameters. This impact needs to be studied and quantified in order for constraints on the BAO shift from simulations to be used to correct measurements from observations.

While all three of these efforts have made significant progress on this important source of systematic uncertainty, there are also limitations to all three studies that make it difficult to identify any one as the definitive measurement of the shift. Furthermore, the three results have not converged on either the exact magnitude nor on the direction of the shift, and all three only detect the shift at about the $\sim3\sigma$ level. In light of this discussion, we decided to take the approach of adding an extra systematic uncertainty to account for a possible BAO shift due to non-linear evolution, rather than estimate the value of the shift and apply it as a correction. 

We add this theoretical systematic component to our total error budget via the covariance matrix of the two BAO parameters:
\begin{equation}
    C(\aper,\apar)_{\rm tot} = C(\aper,\apar)_{\rm stat} + C(\aper,\apar)_{\rm sys}, 
\end{equation}
with
\begin{equation}
C(\aper,\apar)_{\rm sys} = 
\begin{bmatrix}
\Delta \aper^2 & 0 \\
0 & \Delta \apar ^2
\end{bmatrix}
\end{equation}
and 
\begin{eqnarray}
    \Delta \aper &= 0.3\%, \\
    \Delta \apar &= 0.3\%.
\end{eqnarray}
These values are slightly larger than the shifts measured by \cite{deBelsunce2024} ($\sim0.2\%$ for the auto-correlation and $\sim0.1\%$ for the cross-correlation when interpolated to $z=2.33$) and smaller than the 1\% isotropic shift from \cite{Sinigaglia2024}. Given the significant recent interest in the \lya\ BAO shift, there is a high likelihood that much better measurements of this shift will be performed in the near future, such as by applying the EFT approach to real data. We therefore highlight both the statistical and statistical$+$systematic constraints separately to make it easy to update our results in light of this expected future work.

\subsection{DESI DR2 BAO} \label{sec:y3bal}

Our measurements of $\apar$ and $\aper$ from \cref{sec:measurement} combined with the fiducial cosmology lead to the following measurements of the ratios of $D_H$ and $D_M$ relative to $r_d$ at $\zeff=2.33$: 
\begin{equation}
\left\{
\begin{array}{ll}
D_H(\zeff) / r_d = 8.632\pm 0.098 \mathrm{(stat)} \pm 0.026 \mathrm{(sys)}\\
D_M(\zeff) / r_d = 38.99\pm 0.52 \mathrm{(stat)} \pm 0.12 \mathrm{(sys)}\\
\rho(D_H/r_d,D_M/r_d) = -0.457 \mathrm{(stat)}\\
\end{array}
\right.
\end{equation}

Combining both statistical and systematic uncertainties, we obtain our final result: 
\begin{equation}
\left\{
\begin{array}{ll}
D_H(\zeff) / r_d = 8.632 \pm 0.101 \mathrm{(stat+sys)}  \\
D_M(\zeff) / r_d = 38.99 \pm 0.53 \mathrm{(stat+sys)} \\
\rho(D_H/r_d,D_M/r_d) = -0.431 \mathrm{(stat+sys)} \\
\end{array}
\right.
\end{equation}
These are the values used in our companion paper \citep{DESI.DR2.BAO.cosmo}. We suggest the use of these values and uncertainties for testing cosmological models by other studies as well. 

Another common parameter of interest in BAO studies is the isotropic dilation parameter
\begin{eqnarray}
D_V(\zeff)/ r_d &\equiv& \left(\zeff D_M^2 D_H \right)^{1/3}/ r_d \nonumber \\
&=& 31.27 \pm 0.25 \mathrm{(stat)} \pm 0.07 \mathrm{(sys)}~,
\end{eqnarray}
and the anisotropic \citep[or Alcock-Paczy\'nski;][]{AP1979} parameter $f_{\rm AP}\equiv D_M / D_H$:
\begin{equation}
f_{\rm AP}(\zeff) = 4.518 \pm 0.095 \mathrm{(stat)}\pm 0.019 \mathrm{(sys)}~.
\end{equation}
We note that the ratio $D_V / r_d$ is only the optimal definition of the isotropic BAO parameter in the absence of redshift space distortions. Different BAO measurements have different combinations of $D_H$ and $D_M$ that minimize the correlation with $f_{\rm AP}$ and will therefore have a smaller relative uncertainty. The optimal combination for DR2 \lya is approximately 
\begin{equation}
(D_H^{0.55} D_M^{0.45})(\zeff)/ r_d = 17.01 \pm 0.11 \mathrm{(stat)} \pm 0.04 \mathrm{(sys)}.
\end{equation}
This corresponds to a 0.64\% measurement of the isotropic BAO scale at $\zeff=2.33$, or an 0.7\% measurement with the inclusion of the systematic uncertainty.

\begin{figure}
\centering
\includegraphics[width=0.95\columnwidth]{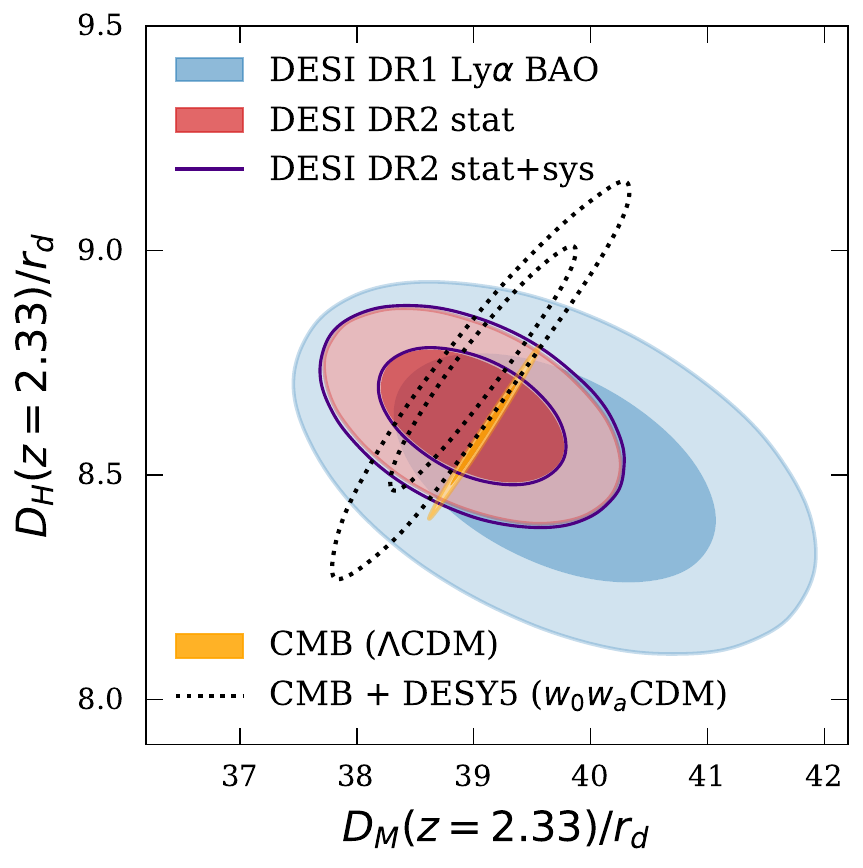}
\caption{\lya BAO measurement of $D_M/r_d$ vs.\ $D_H/r_d$ from DESI DR2 with statistical-only uncertainties (red contour), and statistical$+$systematic uncertainties (solid indigo contour). We compare our result with the previous DESI DR1 measurement (light-blue contour), and constraints inferred from the Planck CMB data assuming $\Lambda$CDM (orange contour), and the combination of Planck CMB and the DES Year 5 supernovae sample assuming $w_0w_a$CDM (dotted black countour). 
}
\label{fig:desi_bao_dmdh} 
\end{figure}

\subsection{Comparison to Previous Work} \label{sec:comp}

We show our measurement of $D_H/r_d$ and $D_M/r_d$ in \Cref{fig:desi_bao_dmdh}, which includes both the statistical-only result (red), and the statistical$+$systematic constraint (indigo). The figure shows the improvement in our BAO constraints between DESI DR1 (blue) and the DR2 results presented here. We also plot the derived constraints on $D_H/r_d$ and $D_M/r_d$ from two cosmological chains using external data. The orange contour shows the Planck cosmic microwave background (CMB) constraint assuming flat $\Lambda$CDM \cite{Planck2018} and the dotted black contour shows the joint Planck CMB and Dark Energy Survey (DES) Year 5 (Y5) supernovae constraints assuming a $w_0w_a$CDM cosmology \cite{DESY5SN:2024}. \Cref{fig:desi_bao_dmdh} shows that our measurement is in good agreement with both of these results, and provides a significant, complementary constraint. This is explored further in the companion paper \citep{DESI.DR2.BAO.cosmo}, which combines this \lya result with BAO measurements from galaxies and lower-redshift quasars.

Our new measurement from DR2 is also in very good agreement with our measurement from DR1 \citep{DESI2024.IV.KP6}, the eBOSS DR16  result \citep{dMdB2020}, and the fiducial cosmology \citep{Planck2018} listed in \cref{tab:fid_cosmo}.
\cref{fig:desi_bao_alpha} shows a comparison of our result with the previous two \lya measurements. The figure demonstrates the robustness of the BAO measurement with the \lya forest. Measurements with progressively larger datasets have remained statistically consistent, and have passed increasingly stringent tests for systematic errors in the analysis. 

The dataset size has increased substantially between the three measurements. The eBOSS DR16 measurement was based on 210,005 quasars with $z > 2.10$ that were used to measure the \lya auto-correlation function and 341,468 quasars at $z > 1.77$ that were used for the cross-correlation function measurement. The DESI DR1 approximately doubled the sample size with over 420,000 quasars for the auto-correlation measurement and over 700,000 for the cross-correlation, although as most of the quasars analyzed for DR1 had only one observation (see \cref{fig:desi_footprint}), the typical SNR of DR1 was on average somewhat lower than for eBOSS DR16. The DESI DR2 sample is nearly twice the size of the DR1 sample, with over 820,000 quasars at $z>2.09$ used for the auto-correlation function measurement and over 1.2 million at $z>1.77$ that contribute to the cross-correlation measurement. In addition, most of the DR2 quasars have multiple observations.  

There were multiple improvements in the data quality between the SDSS spectrographs \citep{Smee2013} and the DESI instrumentation \citep{DESI2022.KP1.Instr}. One is that the DESI corrector system includes an atmospheric dispersion corrector \citep{Corrector.Miller.2023}, which greatly improves the spectro-photometric calibration. Another is that the DESI spectrographs are substantially more stable both because they are gravity invariant (bench mounted) and are located in a climate-controlled room. The greater stability produces a much more stable point spread function that leads to better sky subtraction, as well as better wavelength calibration that leads to smaller redshift errors. Another difference is that the blue channel of the DESI spectrographs has a somewhat higher range of spectral resolution (2000-3000) than the SDSS spectrographs (1500-2000). 

Lastly, there have been significant improvements to the analysis methodology. One significant change from eBOSS to DESI is that all of the DESI \lya analysis was developed with blinded data. Other notable changes from eBOSS to DESI include, a substantial increase in the number of mock datasets, improvements to the continuum fitting code, re-calibration of the spectra, improvements to the weights on \lya pixels, inclusion of the cross-covariance between the four correlation function measurements, and improved descriptions for contamination by metals and high column density systems. The two most significant changes from DESI DR1 to DESI DR2 are to the calculation of the distortion matrix (see \cref{sec:dm}) and modeling metals (see \cref{sec:mm}). We also have adjusted the range of physical separations we include in our correlation function measurements and have included a theoretical systematic error for the first time in a \lya BAO measurement. 

\section{Conclusions} \label{sec:conclusions}

We report the most precise measurement of the BAO scale with the \lya forest to date. This measurement is based on the first three years of operation of the main DESI survey, and these data will be part of the planned second data release (DR2). The sample is approximately a factor of two larger than the DR1 dataset.
The precision of these measurements is 1.1\% 
along the line of sight ($\alpha_\|$) and 1.3\% in the transverse direction ($\alpha_\perp$). The combination yields a statistical precision of 0.65\% in the isotropic BAO parameter at an effective redshift $\zeff = 2.33.$

This statistical precision, in conjunction with several recent theoretical studies, have motivated us to include a systematic error term in the DESI \lya analysis for the first time. Theoretical investigations have detected a BAO shift based on \lya with approximately $3\sigma$ significance \cite{Sinigaglia2024,deBelsunce2024,Hadzhiyska2025}. These studies use different methods and draw somewhat different conclusions on the size of the shift, so we add a systematic uncertainty of $\Delta \alpha_\| = 0.3\%, \Delta \alpha_\perp = 0.3\%$ to our covariance matrix rather than apply a shift correction to our data. This theoretical systematic increases the uncertainty in the isotropic BAO parameter to 0.70\%. 

These measurements of $\alpha_\|$ and $\alpha_\perp$ correspond to measurements of the ratios $D_H(\zeff) / r_d = 8.632\pm 0.098 \mathrm{(stat)} \pm 0.026 \mathrm{(sys)}$ and $D_M(\zeff) / r_d = 38.99\pm 0.52 \mathrm{(stat)} \pm 0.12 \mathrm{(sys)}$, where $D_H(\zeff)$ is the Hubble distance, $D_M(\zeff)$ is the transverse comoving distance at $\zeff=2.33$, and $r_d$ is the sound horizon at the drag epoch.

This paper is one of two key papers that present the BAO measurements from DESI DR2. The other paper \citep{DESI.DR2.BAO.cosmo} presents the measurement of the clustering of galaxies and quasars at $z < 2.1$ and the cosmological interpretation of the full set of DESI DR2 BAO measurements. That includes the consistency of the DESI DR2 BAO measurements with the $\Lambda$CDM model, the evidence for dynamical dark energy, and new results on the sum of the masses of the three neutrino species. These two papers also have five supporting papers. This \lya analysis has supporting papers that describes how we identify DLAs in the data \cite{Y3.lya-s2.Brodzeller.2025} and the synthetic datasets we constructed to validate our analysis \cite{Y3.lya-s1.Casas.2025}. An additional supporting paper \cite{Y3.clust-s1.Andrade.2025} presents our validation of the $z < 2.1$ BAO measurement with galaxies and quasars. Lastly, there are supporting papers that conduct further explorations of dynamical dark energy models \citep{Y3.cpe-s1.Lodha.2025} and neutrinos \citep{Y3.cpe-s2.Elbers.2025}.

We plan additional analyses of the DESI DR2 dataset that will provide more precise measurements of many cosmological parameters. This will include key papers on the full-shape modeling of the clustering of galaxies and quasars, similar to the studies with the DR1 dataset \cite{DESI2024.V.KP5,DESI2024.VII.KP7B,KP7s1-MG}, and the full-shape modeling of the \lya forest \cite{Cuceu2023a,Cuceu2023b} with eBOSS. There will also be key papers on local measurements of primordial non-Gaussianity, the physical properties of the galaxies and quasars, combinations with lensing data, local measurements with peculiar velocities, the mass distribution of the Milky Way, and the construction of the large-scale structure catalogs. 

\section{Data Availability}
The data used in this analysis will be made public along with Data Release 2 (details in \url{https://data.desi.lbl.gov/doc/releases/}). The data points corresponding to the figures from this paper are available from this \url{https://zenodo.org/records/15690869} Zenodo repository.

\begin{acknowledgments}
    
This material is based upon work supported by the U.S. Department of Energy (DOE), Office of Science, Office of High-Energy Physics, under Contract No. DE–AC02–05CH11231, and by the National Energy Research Scientific Computing Center, a DOE Office of Science User Facility under the same contract. Additional support for DESI was provided by the U.S. National Science Foundation (NSF), Division of Astronomical Sciences under Contract No. AST-0950945 to the NSF’s National Optical-Infrared Astronomy Research Laboratory; the Science and Technology Facilities Council of the United Kingdom; the Gordon and Betty Moore Foundation; the Heising-Simons Foundation; the French Alternative Energies and Atomic Energy Commission (CEA); the National Council of Humanities, Science and Technology of Mexico (CONAHCYT); the Ministry of Science, Innovation and Universities of Spain (MICIU/AEI/10.13039/501100011033), and by the DESI Member Institutions: \url{https://www.desi.lbl.gov/collaborating-institutions}. Any opinions, findings, and conclusions or recommendations expressed in this material are those of the author(s) and do not necessarily reflect the views of the U. S. National Science Foundation, the U. S. Department of Energy, or any of the listed funding agencies.

The authors are honored to be permitted to conduct scientific research on Iolkam Du’ag (Kitt Peak), a mountain with particular significance to the Tohono O’odham Nation.

\end{acknowledgments}

\bibliographystyle{mod-apsrev4-2} 
% \bibliography{main,desi}
%apsrev4-2.bst 2018-12-27 (MD) hand-edited version of apsrev4-1.bst



\appendix

\section{Nuisance Parameters} \label{app:nuisance}

Our fit to the four correlation functions has 17 parameters: the two BAO parameters $\apar$ and $\aper$, and 15 nuisance parameters. \cref{tab:baseline} lists all of these parameters, the priors on each of them, the best-fit values for the combined (baseline) fit, and the best-fit values from fits to just the two auto-correlations and just the two cross-correlations. Note that not all of the nuisance parameters are needed in the fits to just the auto-correlation or just the cross-correlation. The table also lists the number of data points ($N_{bin}$) in each fit, the number of parameters ($N_{param}$), the minimum $\chi^2$ value of the fit ($\chi^2_{min}$), and the fit p-value.  

The first two nuisance parameters in \cref{tab:baseline} are the bias and RSD parameters of the \lya forest, $b_\alpha$ and $\beta_\alpha$, respectively. The next five are bias values for the main metal lines that we expect in the \lya forest region. These are $b_{\rm SiII(1190)}, b_{\rm SiII(1193)}, b_{\rm SiII(1260)}, b_{\rm SiIII(1207)}$, and $b_{\rm CIV(eff)}$, where the last one is labeled as ``eff'' instead of with the rest-frame wavelength in \AA\ because it also includes some contribution from other metal lines at longer wavelengths (especially Mg\,II and Si\,IV). We have three parameters for high column density (HCD) systems: a bias parameter $b_{HCD}$, an RSD parameter $\beta_{HCD}$, and a scale $L_{HCD}$. The additional parameter $L_{HCD}$ is the width of a filter that is used to model the wings of the HCDs. The parameter $b_Q$ is the quasar bias. When we fit the cross-correlation alone, we set a tighter prior on the quasar bias of $b_Q = 3.5 \pm 0.1$ \citep{Chaussidon2024} to break the degeneracy between $b_Q$ and $b_\alpha$. We have two parameters to account for redshift errors:  $\Delta r_\|$, which allows for a systematic shift in quasar redshifts, and $\sigma_v$, which accounts for the combination of quasar peculiar velocities and redshift errors \citep{KP6s4-Bault}. The parameter $\xi_0^{\rm TP}$ sets the amplitude of the proximity effect, which accounts for a combination of the higher radiation field in the vicinity of quasars and the higher gas density \citep{FontRibera2013}. Lastly, the parameter $a_{noise}$ accounts for correlated noise between spectra from fibers in the same DESI focal plane petal, which is dominated by the sky background model \citep{KP6s5-Guy}. We refer to \DESIIV for a more detailed discussion of the nuisance parameters and how they are included in the model. 

The nuisance parameters are generally of the same order as their values in \DESIIV, and agree reasonably well between the best fit combined, auto-only, and cross-only results. We do not expect exact agreement between the nuisance parameters, as their purpose is to approximately account for known physical effects that could impact the goodness-of-fit of the model, and not to accurately measure these effects. When we analyze the auto-correlation or the cross-correlation alone, we fix $L_{\rm HCD}$ to the best-fit value of the combination ($L_{\rm HCD}=5.3~\hMpc$).
This is necessary to break internal degeneracies, but it makes the p-value of these analyses difficult to interpret. Similar to \DESIIV, we do not measure $b_Q$, $\Delta r_\|$, $\sigma_v$, and $\xi_0^{\rm TP}$ when we fit the auto-correlation alone, as these quantities are only relevant to the quasar tracers, and we do not fit $b_{\rm CIV(eff)}$ and $a_{noise}$ when we fit to the cross-correlation alone, as these quantities are only important in the auto-correlation fit. We do not include a model for the impact of the relativistic effects described in \citep{Irsic2016}, as was done in some previous eBOSS analyses \citep{Bautista2017,Blomqvist2019}. This was not included because of degeneracies with other effects that can cause a dipole in the cross-correlation, including metal contamination and redshift errors. While \citet{Irsic2016} showed that it should not impact the location of the BAO peak, \citet{Lepori2020} demonstrated that DESI should have the statistical power to measure this effect, and we plan to include these effects in future work. 

\begin{table*}
\centering
\begin{tabular}{c|c|ccc}
Parameter                                 & Priors                                    & \multicolumn{3}{c}{Best fit}\\
                                          &                                           & Combined                         & Ly$\alpha \times$Ly$\alpha$      & Ly$\alpha \times$QSO             \\
\hline

$\alpha_{\parallel}$                      & $\mathcal{U}[0.01, 2.00]$                 & $1.002 \pm 0.011$                & $1.000 \pm 0.014$                & $1.004 \pm 0.015$                \\
$\alpha_{\perp}$                          & $\mathcal{U}[0.01, 2.00]$                 & $0.995 \pm 0.013$                & $1.003 \pm 0.020$                & $0.985 \pm 0.016$                \\
$b_{\alpha}$                            & $\mathcal{U}[-2.00, 0.00]$                & $-0.1352 \pm 0.0073$             & $-0.1488 \pm 0.0020$             & $-0.099 \pm 0.021$               \\
$\beta_{\alpha}$                        & $\mathcal{U}[0.00, 5.00]$                 & $1.445 \pm 0.064$                & $1.365 \pm 0.032$                & $1.98 \pm 0.35$                  \\
$10^3 b_{\rm SiII(1190)}$                 & $\mathcal{U}[-500.00, 0.00]$              & $-3.70 \pm 0.39$                 & $-3.80 \pm 0.44$                 & $-3.24 \pm 0.74$                 \\
$10^3 b_{\rm SiII(1193)}$                 & $\mathcal{U}[-500.00, 0.00]$              & $-3.18 \pm 0.38$                 & $-3.55 \pm 0.43$                 & $-2.56 \pm 0.72$                 \\
$10^3 b_{\rm SiII(1260)}$                 & $\mathcal{U}[-500.00, 0.00]$              & $-3.67 \pm 0.40$                 & $-3.18 \pm 0.48$                 & $-4.28 \pm 0.62$                 \\
$10^3 b_{\rm SiIII(1207)}$                & $\mathcal{U}[-500.00, 0.00]$              & $-7.3 \pm 1.5$                   & $-12.2 \pm 1.5$                  & $-3.8 \pm 1.9$                   \\
$10^3 b_{\rm CIV(eff)}$                   & $\mathcal{N}(-19.0, 5.0)$*                & $-18.6 \pm 4.9$                  & $-19.6 \pm 5.0$                  &                                  \\
$b_{HCD}$                                 & $\mathcal{U}[-0.20, 0.00]$                & $-0.0206 \pm 0.0090$             & $-0.000 \pm 0.017$               & $-0.060 \pm 0.022$               \\
$\beta_{HCD}$                             & $\mathcal{N}(0.500, 0.090)$               & $0.508 \pm 0.089$                & $0.500 \pm 0.090$                & $0.510 \pm 0.090$                \\
$L_{\rm HCD} (h^{-1} {\rm Mpc})$          & $\mathcal{N}(5.0, 1.0)$                   & $5.30 \pm 0.93$                  & $5.00 \pm 1.00$                  & $4.9 \pm 1.0$                    \\
$b_{Q}$                                 & $\mathcal{N}(3.40, 0.20)$                 & $3.545 \pm 0.054$                &                                  & $3.41 \pm 0.19$                  \\
$\Delta r_{\parallel} (h^{-1} {\rm Mpc})$ & $\mathcal{N}(0.0, 1.0)$                   & $0.53 \pm 0.18$                  &                                  & $0.72 \pm 0.20$                  \\
$\sigma_v (h^{-1} {\rm Mpc})$             & $\mathcal{U}[0.00, 15.00]$                & $3.18 \pm 0.64$                  &                                  & $3.5 \pm 1.1$                    \\
$\xi_0^{\rm TP}$                          & $\mathcal{U}[0.00, 2.00]$                 & $0.453 \pm 0.046$                &                                  & $0.349 \pm 0.071$                \\
$10^4 a_{\rm noise}$                      & $\mathcal{U}[0.00, 100.00]$               & $2.20 \pm 0.15$                  & $2.19 \pm 0.15$                  &                                  \\
\hline
$N_{\rm bin}$                             & --                                        & 9306                             & 3102                             & 6204                             \\
$N_{\rm param}$                           & --                                        & 17                               & 13                               & 15                               \\
$\chi^2_{\rm min}$                        & --                                        & 9304.46                          & 3146.82                          & 6133.10                          \\
p-value                                   & --                                        & 0.45                             & 0.23                             & 0.69                             \\
\end{tabular}
\caption{Priors, best-fit values (mean of the posterior) and uncertainties (68\% credible intervals) for the 17 free parameters in the fits. Some parameters are not needed when fitting the auto-correlation or the cross-correlation alone.
}
\label{tab:baseline}
\end{table*}

\section{Data Validation} \label{app:datavalidation}

\begin{figure*}
\centering
\includegraphics[width=0.79\textwidth]{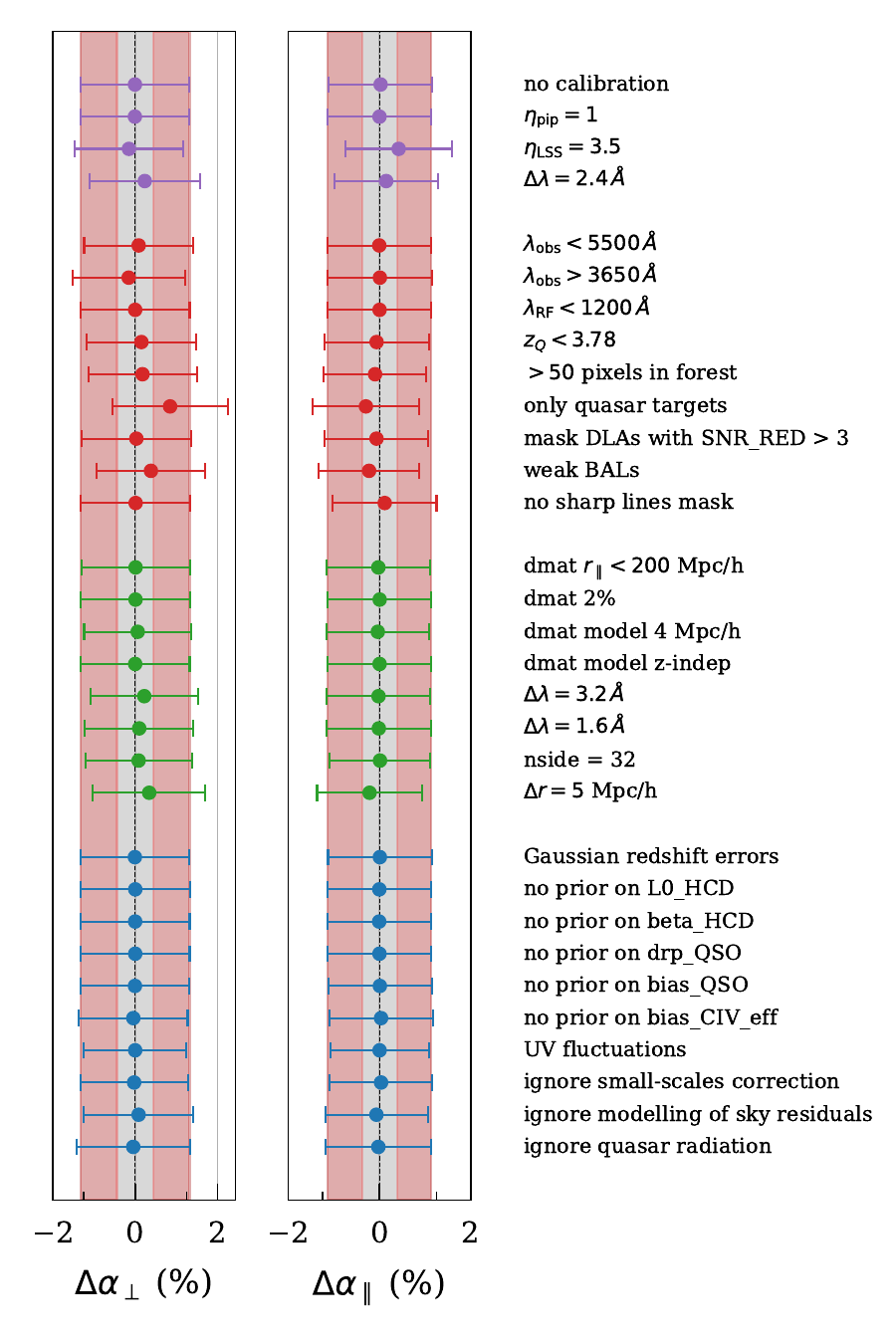}
\caption{One-dimensional shifts in the BAO parameters for alternative analysis choices. These include variations in the method to estimate the fluctuations (purple), the dataset (red), the method to compute correlations and covariances (green), and modeling choices (blue). The red shaded regions show the one $\sigma$ uncertainty from the main analysis and the smaller gray area shows the threshold set to these tests ($\sigma / 3$). The two parameters are anti-correlated with $\rho=-0.48$.
The only shift that exceeds the $\sigma/3$ threshold is the ``only quasar targets'' dataset variation. This larger shift is consistent with the change in the sample size. See \cref{app:datavalidation} for more details.
}
\label{fig:var_all_1d}
\end{figure*}

This appendix describes additional data analysis variations beyond those already described in \cref{sec:dataval}. All were run as part of the tests before unblinding our analysis, and all were rerun after unblinding on the final DR2 dataset (\texttt{loa}). The results are summarized in \cref{fig:var_all_1d} and all but one of these shifts pass our $\sigma/3$ criterion. The one apparent exception is the ``only quasar targets'' variation. This variation corresponds to a more significant sample size change than the other variations. We confirmed that the size of the shift for this variation is consistent with the sample size difference relative to the baseline sample. These variations are discussed in greater detail in \DESIIV. 

The tests are split into four categories based on the nature of the analysis variation. The first set with violet errorbars show the BAO shifts from alternative analyses with a different estimation of \lya\ forest fluctuations, without changing the dataset. The variations are as follows: 1) \texttt{no calibration}: We do not perform the recalibration of the spectra using the C\,{\sc III} region described in \cite{Ramirez2024}; \texttt{$\eta_{\rm pip} = 1$}: We do not apply the recalibration of the instrumental noise $\eta$ that is used in the baseline analysis; \texttt{$\eta_{\rm LSS} = 3.5$}: We use non-optimal weights and reduce by a factor of two the contribution from the intrinsic \lya forest variance to the weights; \texttt{$\Delta \lambda = 2.4$\,\AA}: We coadd three pixels into one before performing the continuum fitting and assigning weights. In this case we use the value of $\eta_{\rm LSS} = 3.1$ that was found to be optimal for this coarser pixelization in \cite{Ramirez2024}.

The red errorbars show the BAO shifts from alternative analyses that result in small changes of the dataset. These are: \texttt{$\lambda_{\rm obs} < 5500$\,\AA}: We only use \lya\ pixels below this observed wavelength, rather than the baseline value of $\lambda_{\rm obs} < 5577$ \AA; \texttt{$\lambda_{\rm obs} > 3650$\,\AA}: We only use \lya\ pixels above this observed wavelength, rather than the baseline value of $\lambda_{\rm obs} > 3600$\,\AA; \texttt{$\lambda_{\rm RF} < 1200$\,\AA}: We only use \lya\ pixels below this rest-frame wavelength, rather than the baseline value of $\lambda_{\rm RF} < 1205$ \AA. \texttt{$z_{\rm QSO} < 3.78$}: We only include quasars with $z_{\rm QSO}=3.78$, which is the highest redshift included in the mocks; \texttt{$> 50$ pixels in forest}: We include lines-of-sight with more than 50 valid \lya\ pixels. For the baseline analysis we require at least 150 pixels; \texttt{only quasar targets}: We only use quasars that were considered quasar targets. This variation causes a more significant change in the size of the dataset as described above; \texttt{Mask DLAs in spectra with ${\rm SNR\_RED}>3$ DLAs}: We only mask DLAs identified in quasar spectra with ${\rm SNR\_REDSIDE}>3$, instead of the ${\rm SNR\_REDSIDE}>2$ threshold used in the baseline analysis; \texttt{Weak BALs}: We do not include the \lya\ forest of those quasars where we have identified very strong BAL features. This only excludes BALs with $AI > 840$, corresponding to the 50\% percentile of strongest BALs  \cite{KP6s9-Martini}. \texttt{No sharp lines mask}: We do not mask the four sharp lines discussed in \cite{Ramirez2024}, related to sky lines and Calcium absorption features from the interstellar medium of the Milky Way.

The green errorbars show the BAO shifts from alternative measurements of the correlation functions, their covariances, and the distortion matrices. We look at the following variations: \texttt{dmat $r_\parallel < 200$ Mpc/h}: We only model the distortion matrix up to $r_\parallel = 200$ Mpc/h, while the baseline analysis extends the model to $r_\parallel = 300$ Mpc/h; \texttt{dmat 2\%}: We use 2\% of the dataset to compute the distortion matrix, rather than 1\% as in the baseline analysis; \texttt{dmat $dr = 4$ Mpc/h}: We model the distortion matrix using the same binning as in the measurement of the correlation function (4 Mpc/h), rather than 2\,Mpc/h binning  as in the baseline analysis; \texttt{dmat model z-indep}: We ignore the redshift evolution in the computation of the distortion matrix, as was done in eBOSS and in DESI-DR1; \texttt{$\Delta \lambda = 3.2$\,\AA}: We re-bin the continuum-fitted deltas by four pixels (rebinned pixels of 3.2\,\AA), rather than by three pixels as in the baseline analysis (rebinned pixels of 2.4\,\AA); \texttt{$\Delta \lambda = 1.6$ \AA}: We re-bin the continuum-fitted deltas by two pixels (rebinned pixels of 1.6\,\AA), rather than by three pixels as in the baseline analysis; \texttt{nside $= 32$}: We measure the correlations in HEALPix pixels defined by nside=32 (instead of nside=16 in the baseline), which results in 3646 HEALPix pixels with at least one quasar (rather than of 1028 HEALPix pixels in the baseline analysis); 
\texttt{$\Delta r = 5$ Mpc/h}: We use a 5 Mpc/h binning of the correlation function, rather than 4\,Mpc/h as in the baseline analysis.

The blue errorbars show the BAO shifts from alternative analyses with different modeling choices. We investigate the following variations:
\texttt{Gaussian redshift errors}: We use a Gaussian distribution to model quasar redshift errors and quasar peculiar velocities, rather than the Lorentzian model used in the baseline analysis;
\texttt{no prior on L0\_HCD}: We remove the informative prior on the parameter $L_{\rm HCD}$ in the model of the contamination by HCDs;
\texttt{no prior on beta\_HCD}: We remove the informative prior on the parameter $\beta_{\rm HCD}$ in the model of the contamination by HCDs;
\texttt{no prior on drp\_QSO}: We remove the informative prior on the parameter $\Delta r_\parallel$ in the model of quasar redshift errors;
\texttt{no prior on bias\_QSO}: We remove the informative prior on the quasar bias parameter $b_Q$; 
\texttt{no prior on bias\_CIV\_eff}: We remove the informative prior on the parameter $b_{\rm CIV, eff}$ in the model of the contamination by CIV absorption;
\texttt{UV fluctuations}: We model the impact of fluctuations in the UV background \cite{Pontzen2014,Gontcho2014} on the \lya forest auto-correlation following the prescription of \cite{Bautista2017};
\texttt{ignore small-scales correction}: We ignore the small-scale, multiplicative correction from \cite{Arinyo2015} in the model of the \lya auto-correlation;
\texttt{ignore modeling of sky residuals}: We ignore the contamination from correlated sky residuals in the \lya auto-correlation, discussed in \cite{KP6s5-Guy};
\texttt{ignore quasar radiation}: We ignore the transverse proximity effect, the impact of quasar radiation in the cross-correlation. The $\chi^2$ and $\Delta \chi^2$ values (relative to the baseline) of these different modeling variations are listed in \cref{tab:chisq}. The $\Delta \chi^2$ values are significant for the last three, all of which remove effects that we expect to be present in the data. We include these variations to demonstrate our implementation does not impact BAO.  

\begin{table*}
\centering
\begin{tabular}{lcccr}
Variation                                 & $\chi^2$ & $N_{bin}$ & $N_{param}$ & $\Delta \chi^2$ \\ 
\hline
Baseline                            & 9304.46 &  9306 & 17 &    \\
Gaussian redshift errors            & 9303.77 &  9306 & 17 &  -0.69 \\
no prior on L0\_hcd                  & 9303.71 &  9306 & 17 &  -0.75 \\
no prior on beta\_hcd                & 9303.80 &  9306 & 17 &  -0.66 \\
no prior on drp\_QSO                 & 9304.17 &  9306 & 17 &  -0.29 \\
no prior on bias\_QSO                & 9303.90 &  9306 & 17 &  -0.56 \\
no prior on bias\_CIV\_eff            & 9304.22 &  9306 & 17 &  -0.24 \\
UV fluctuations                     & 9304.45 &  9306 & 18 &  -0.01 \\
ignore small-scales correction      & 9296.87 &  9306 & 17 &  -7.59 \\
ignore modelling of sky residuals   & 9505.22 &  9306 & 16 & 200.76 \\
ignore quasar radiation             & 9409.91 &  9306 & 16 & 105.45 \\
\end{tabular}
\caption{Change in $\chi^2$ for the modeling variations shown at the bottom of  \cref{fig:var_all_1d}. For each variation listed in column one we list the $\chi^2$ value, the number of bins, the number of parameters, and the difference in $\chi^2$ relative to the baseline (variation -- baseline). See \cref{sec:dataval} for discussion of these values. 
}
\label{tab:chisq}
\end{table*}

\end{document}